\documentclass[a4paper,fleqn,usenatbib,useAMS]{mnras}

\usepackage{graphicx}	
\usepackage{amsmath}	
\usepackage{amssymb}	
\usepackage{multicol}        
\usepackage{bm}		
\usepackage{pdflscape}	
\usepackage{longtable}	

\usepackage[T1]{fontenc}
\usepackage{ae,aecompl}

\usepackage{mathptmx}

\title[New insights on the dissociative merging galaxy cluster Abell 2034]{New insights on the dissociative merging galaxy cluster Abell 2034}

\author[Monteiro-Oliveira et al.] 
  {R.~Monteiro-Oliveira,$^{1,2}$\thanks{E-mail: rogerionline@gmail.com}
  E. S.~Cypriano,$^1$
  A. Z.~Vitorelli,$^1$
  A. L. B.~Ribeiro,$^3$
  \newauthor 
  L.~Sodr\'e Jr.,$^1$
  R.~Dupke$^{4,5,6,7}$
  and
  C.~Mendes de Oliveira$^1$\\
  $^1$Universidade de S\~ao Paulo, Inst. de Astronomia, Geof\'isica e Ci\^encias Atmosf\'ericas, Depto. de Astronomia, R. do Mat\~ao 1226, 05508-090 S\~ao Paulo, Brazil\\
  $^2$Universidade Federal do Rio Grande do Sul, Instituto de F\'isica, Departamento de Astronomia, Campus do Vale, 91501-970, Porto Alegre, Brazil\\
  $^3$Universidade Estadual de Santa Cruz, Laborat\'orio de Astrof\'isica Te\'orica e Observacional - 45650-000, Ilh\'eus-BA, Brazil\\
  $^4$Observat\'orio Nacional,  Rua Gal. Jos\'e Cristino 77, 20921-400 Rio de Janeiro RJ, Brazil\\
  $^5$University of Michigan, Department of Astronomy, 311 West Hall 1085 South University Ave. Ann Arbor, MI 48109-1107 USA\\
  $^6$University of Alabama, Department of Physics and Astronomy, Box 870324, Tuscaloosa, AL 35487, USA\\
  $^7$Eureka Scientific Inc., 2452 Delmer St. Suite 100, Oakland, CA 94602, USA
    }

\date{Accepted 2018 August 24. Received 2018 July 30; in original form 2017 December 31}

\pubyear{2018}

\begin{document}
\label{firstpage}
\pagerange{\pageref{firstpage}--\pageref{lastpage}}
\maketitle

\begin{abstract}
We present here new insights about the merging galaxy cluster Abell 2034 ($\bar{z}=0.114$) based on a combined weak lensing and dynamical analysis. From our deep Subaru $BR_Cz'$ images plus Gemini-GMOS/N low-resolution spectra accompanied by available redshift data, we have obtained the individual masses of the colliding subclusters as well as estimated a timeline of the process. The collision event happened $0.56^{+0.15}_{-0.22}$ Gyr ago between the subclusters A2034S ($M_{200}^S=2.35_{-0.99}^{+0.84}\times 10^{14}$ M$_{\odot}$) and A2034N ($M_{200}^N=1.08_{-0.71}^{+0.51}\times 10^{14}$~M$_{\odot}$) with the gas content of both subclusters displaced in relation to their galaxy and dark matter distributions, in a scenario similar to that found in the Bullet Cluster.  Following our data and modelling the collision event is, most likely, taking place not so far from the plane of the sky, with an angle of  $27^\circ\pm14^\circ$ in relation to that. In spite of the intrinsic degeneracy inherent to the system (whether it  has been observed incoming or outgoing), the comparison of our calculated time since the closest approximation with the estimated age of the observed X-ray shock front and the increment experienced by the velocity dispersion of the galaxy cluster members points toward an outgoing movement. Besides, we found a strong evidence for the presence of a third structure which we called A2034W.

\end{abstract}




\section{Introduction}

Immediately after the equipartition epoch, matter became the dominant species relative to the radiation content of the Universe. Baryonic matter and radiation were strongly coupled preventing the growth of small baryonic matter overdensities. Dark matter (DM), on the other hand was not significantly coupled to this fluid, allowing their collapse and posterior growing to form more  massive structures. After recombination, baryons started falling into the potential wells created by DM halos that were formed. From that period and on, structures have evolved due to the gravitational attraction. In the hierarchical (or bottom-up) scenario this is followed by a continuously ongoing process where smaller DM halos merge to form the large galaxy clusters we now observe, the largest gravitationally bounded structures in the Universe. Some of these mergers have reached energies not seen since the Big Bang \citep[$\approx 10^{64}$ ergs;][]{sarazin04}.

Merging galaxy clusters have proved to be fruitful laboratories to study their three main components (DM, galaxies and the hot intra-cluster gas -- ICM) as well the interactions among them. Especially interesting are the dissociative mergers, where systems are observed with significant spatial detachment between the DM and ICM distribution \citep[e.g.][]{Monteiro-Oliveira17a,Monteiro-Oliveira17b}.

\begin{figure*}
 \begin{center}
\includegraphics[width=\textwidth, angle=0]{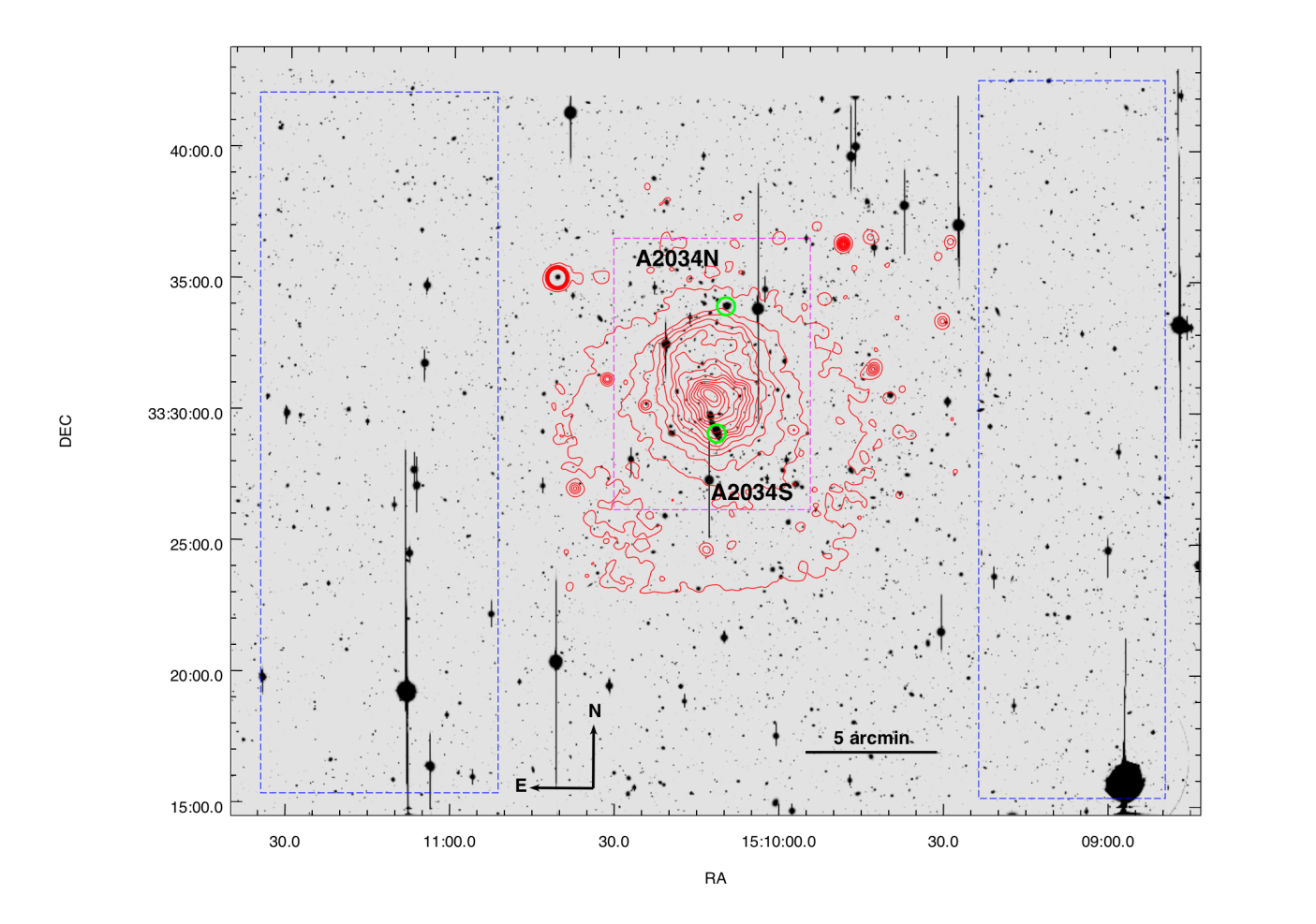}
\caption{Deep optical $R_C$ image of the A2034 field taken by the Suprime-Cam mounted on the Subaru telescope. The overlaid red contours correspond to the ICM X-ray emission mapped by {\it Chandra} telescope. The X-ray emission presents a single peak located between the two BCGs (green circles). The explanation for the south X-ray emission excess is still an open question. The boxes enclose the regions considered for the statistical subtraction aiming to identify the galaxies belonging to the cluster red sequence: the blue ones are sufficiently far from the cluster centre (box {\it magenta}) and are supposed to be dominated by field galaxies.} 
\label{fig:A2034.field}
\end{center}
\end{figure*}

The galaxy cluster Abell 2034 (A2034) was classified by \cite{abell58} as a richness 2 system. It has been exhaustively observed by ROSAT \citep{davidformanjones99}, ASCA \citep{white00}, XMM-{\it Newton} \citep{sevenmergers} and {\it Chandra} \citep{kempner03, owers14}, whose data revealed that A2034 is, indeed, a bimodal system comprised by two subclusters each one located at South and North of the field. In this work, we name them as A2034S and A2034N, respectively. The cluster field is presented in Fig.~\ref{fig:A2034.field}.

The subclusters were identified as galaxy overdensities near each brightest cluster galaxy (BGC), respectively BCG S and BCG N, which are $\sim$ 5 arcmin apart from each other. The X-ray morphology presents itself as a unimodal distribution, offcentred $91\pm1$ arcsec (this work) from the nearest BCG S, suggesting that A2034 as a whole is out of equilibrium. Moreover, \cite{kempner03} found an  discontinuity in the X-ray surface brightness, distant $\sim$3 arcmin from the BCG S toward BCG N, that they classified as a cold front. In the same region, \cite{kempnersarazin01} found an elongated radio emission. All these evidences points towards a recent merger between A2034S and A2034N.

An apparent contradiction to the post merger scenario was the detection by \cite{white00} of a high cooling flow rate (about $\sim 90-580$ M$\odot$ year$^{-1}$), a feature that is correlated with undisturbed systems. However, \cite{kempner03} recalculated this value to $23^{+21}_{-20}$ M$\odot$ year$^{-1}$ and argued that the previous high value was biased by the cold front and the cold gas found at the excess emission region in the Southern of the cluster. The cluster  X-ray luminosity is also high, at $L_{[0.1-2.4~{\rm keV}]}=3.51\times10^{44}$ erg s$^{-1}$ \citep{Piffaretti11}.

The excess emission region is made of a gas colder than that found elsewhere in the nearby field. \cite{kempner03} argued that its temperature is sufficiently low to be a post collisional gas. They also speculated that its origin could be a background galaxy cluster at $0.30 < z< 1.25$ according to the $L_X-T$ relation. For the whole cluster, \cite{davidformanjones99} found $L_{[0.5-2{~\rm keV}]}=5.2\times 10^{44}$ erg and an X-ray bolometric luminosity of $2.2\times 10^{45}$ erg s$^{-1}$.

Based on $1.4 ~\mathrm{GHz}$ VLA data, \cite{giovannini09} confirmed the existence of an radio diffuse emission previously identified by \cite{davidformanjones99}. However, \cite{vanweeren11} revisited this cluster and questioned this classification since the source does not show a clear correlation with the X-ray distribution. They have suggested that the radio source could be a radio relic, but they could not agree to a definitive classification. More recently, \cite{Shimwell16} have presented a complete radio study of the field with Low-Frequency Array (LOFAR) data. They found that the ICM presents a complicated emission and puts A2034 as a more complex system than expected. The brightest X-ray region also coincides with a giant radio halo and a group of three radio relics candidates were also detected in the cluster neighbourhoods

\cite{sevenmergers} performed mass reconstruction through weak lensing using Subaru $R_C$ (2800 s) and $g'$ (720 s) images. They found a more complex scenario than that revealed by the X-ray observations. They found mass counterparts for the subclusters A2034S and A2034N (placed in front of the cold front) and other ones to the west from the cluster. Based on SDSS spectroscopic data, they speculated that them could be background structures. In Tab.~\ref{tab:mass.comp} we show a compilation of the mass estimates for A2034 available in the literature.

\begin{table}
\begin{center}
\caption[]{Compilation of the mass estimates for A2034 available in the literature.}
 \begin{tabular}{c c c}
 \hline \hline 
M$_{200}$ & Method & Reference \\
($10^{14}$ M$_{\odot}$) $h_{70}^{-1}$ & & \\[5pt]
\hline 
$5.00\pm0.03$    & Caustic    & \cite{geller13}\\[5pt]
$11\pm4$ 		 & Caustic    & \cite{owers14}\\[5pt]
$4.30_{+4.00}^{-2.40}$  & WGL$^1$-NFW     & \cite{delliou15}$^2$\\[5pt]
$3.54_{-1.17}^{+0.95}$  & WGL - NFW   & This work (Sec.~\ref{sec:wl})\\
\hline \hline 
\multicolumn{3}{l}{$^1$ Weak gravitational lensing.}\\
\multicolumn{3}{l}{$^2$ Based on M$_{\rm vir}=7.17\pm4.3\times 10^{14}$ M$_{\odot}$  $h_{70}^{-1}$ obtained by}\\
\multicolumn{3}{l}{\cite{sevenmergers}}\\
\end{tabular}
\label{tab:mass.comp}
\end{center}
\end{table}

Using deep {\it Chandra} images and radial velocities of the member galaxies, \cite{owers14} claimed that A2034 is, in fact, a post collisional system whose axis is nearly coincident with the N-S direction. They re-classified the X-ray feature behind A2034N as a shock front travelling with $v_{\rm shock}\simeq2057$ km s$^{-1}$ (corresponding to a Mach number $\mathcal{M}=1.59_{-0.07}^{+0.06}$). The gas content of A2034N then would have been ripped out due to the collision, which happened 0.3 Gyr ago and had a small impact parameter, as revealed by the X-ray morphology. Combining the radial velocities of the BCGs, they estimate that the collision axis is at an angle of $\sim 23^\circ$ respective to the plane of the sky. Moreover, they have found an absence of background galaxies in deep SDSS images which exclude the idea of an background cluster responsible for the X-ray excess emission as suggested by \cite{kempner03}. They have interpreted this feature as a gas lost by A2034N during the passage though A2034S.

In this paper we intend to map the mass distribution in A2034 as well to measure the masses of the individual subclusters. We also plan to characterize the system through the dynamical point of view, aiming to recover the merger history of the system. To these purposes, we have used deep $B$, $R_c$ and $z'$ images obtained with the Subaru telescope and spectroscopic data taken with Gemini/N telescope complemented with data available in the literature.

This paper is organized as follows. In Section~\ref{sec:wl} we present the weak lensing analysis, from the description of the data to the mass measurements.  The dynamical overview based on the galaxy redshift analysis can be found in Section~\ref{sec:dynamical}. The proposed merger scenario for A2034 is described in Section~\ref{sec:two.body}. All results obtained are discussed in Section \ref{sec:discussion} and summarized in Section \ref{sec:summary}.

Throughout this paper we adopt the following cosmology: $\Omega_m=0.27$, $\Omega_\Lambda=0.73$, $\Omega_k=0$, and $H_0=70$~km~s$^{-1}$ Mpc$^{-1}$.  At the mean cluster redshift of $z=0.114$ we then have $1$ arcsec equals $2.06$~kpc, the age of the Universe $12.4$~Gyr, and an angular diameter distance of $424.6$~Mpc.

\section{Weak lensing analysis}
\label{sec:wl}

For the sake of conciseness, we refer the reader interested in the basic concepts of the weak lensing technique to our previous works \citep{Monteiro-Oliveira17a,Monteiro-Oliveira17b}. Complementary, a more complete and detailed description of the subject can be found, for example, at  \cite{mellier99}, \cite{schneider05} and \cite{schneider06}.

\subsection{Imaging observation and reduction}
\label{sec:ior}

Our deep $B$, $R_C$ and $z'$ imaging data were taken by the Suprime-Cam mounted at Subaru Telescope within the Gemini Telescope time exchange program (GN-2007A-C-21). The observations, summarized in the Tab.~\ref{fig:imaging} were done in queue mode in 2007A, mostly under photometric conditions.

\begin{table}
\begin{center}
\caption{Imaging characteristics.}
\begin{tabular}{lccc}
\hline
\hline
Band & Total exposure (h) & Seeing (arcsec) & Completeness\footnotemark[1] \\
\hline
$B$   & 1.50 & 0.98 & 26.6 \\ 
$R_C$ & 2.78 & 1.21 & 26.5  \\
$z'$  & 1.02 & 0.91 & 25.6 \\
\hline
\hline
\multicolumn{4}{l}{$^1$ Corresponds to the magnitude in which the logarithmic}\\
\multicolumn{4}{l}{counts drop 5\% by comparison to the Subaru Deep Field }\\
\multicolumn{4}{l}{\citep{sdf}.}
\end{tabular}
\label{fig:imaging}
\end{center}
\end{table}

Imaging data reduction was performed with a tailor-made software {\sc SDFRED} \citep[][]{sdfred1,sdfred2}. This semi-automatic routine include nine basic steps: ($i$) bias and overscan subtraction, ($ii$) flatfielding, ($iii$) atmospheric and dispersion corrections, ($iv$) sky subtraction, ($v$) auto guide masking, ($vi$) alignment (done for all three filters simultaneously) ($vii$) combining and mosaicing into a final image per filter, ($viii$) fringing removal from $z'$ images, and ($ix$) registering and combining the images (with IRAF).

We have used standard star fields \citep[][]{landolt92,smith02} to calibrate magnitudes in the AB system and perform the astrometric calibration. The object catalogues were built with {\sc SExtractor} \citep{sextractor} in ``double image mode'', where the detections were always made in the $R_C$ band. Galaxies were selected according to two complementary criteria: for $19 \leq R_C \leq 26.5$ galaxies were the objects with ${\rm FWHM} > 1.53$ arcsec and for the brightest objects ($R_C < 19$), galaxies were identified as having {\sc SExtrator}'s ${\rm CLASS\_STAR} < 0.8$. Stars (actually point sources) were selected by their stellarity index (bright saturated stars) or FWHM. For our forthcoming analysis, we have considered as the weak lensing image our deepest $R_C$ band.

\subsection{Classifying the galaxy populations}
\label{sec:member}

For weak lensing purposes we are interested only in the background galaxies. However, we need to carefully exclude from the sample both cluster members and foreground galaxies in order to avoid the introduction of additional noise in the lens model. Besides, the unambiguous identification of the red cluster members is important in order to compare their spatial distribution with the other cluster components (dark matter and hot X-ray emitting gas).

\begin{figure*}
\begin{center}
\includegraphics[width=\columnwidth, angle=0]{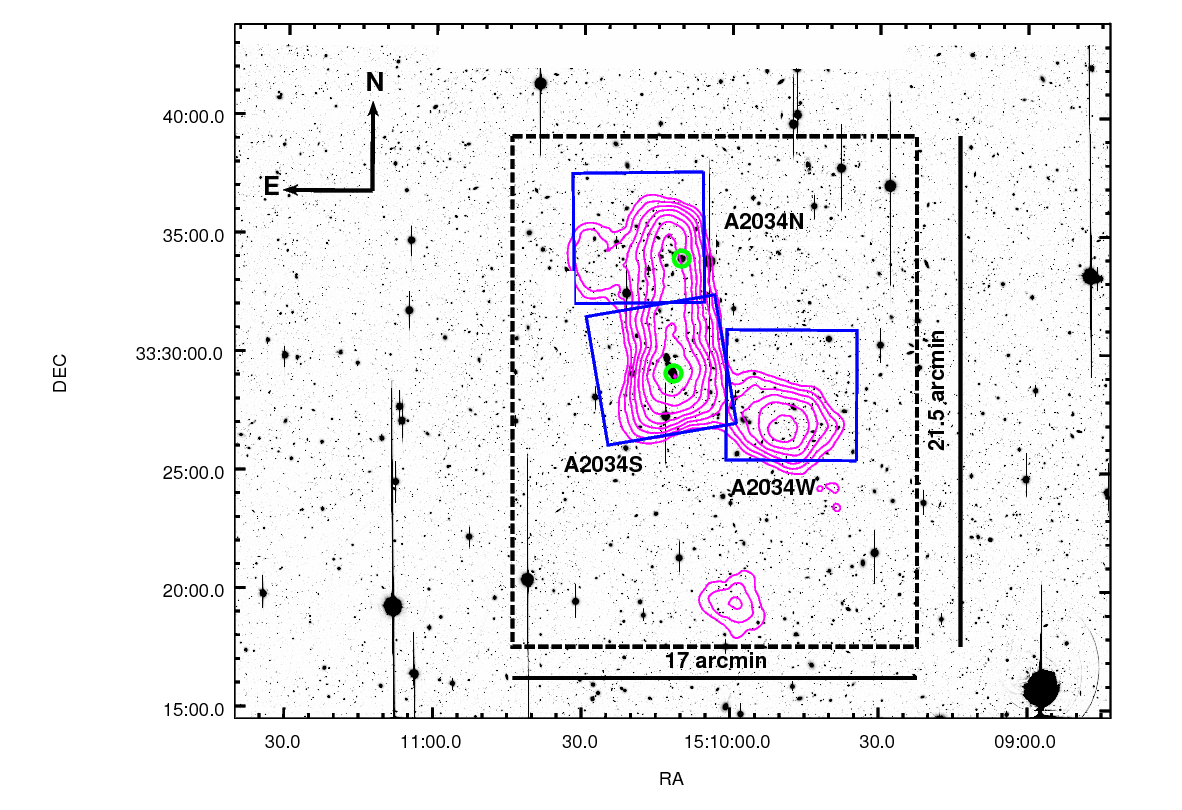}\quad
\includegraphics[width=\columnwidth, angle=0]{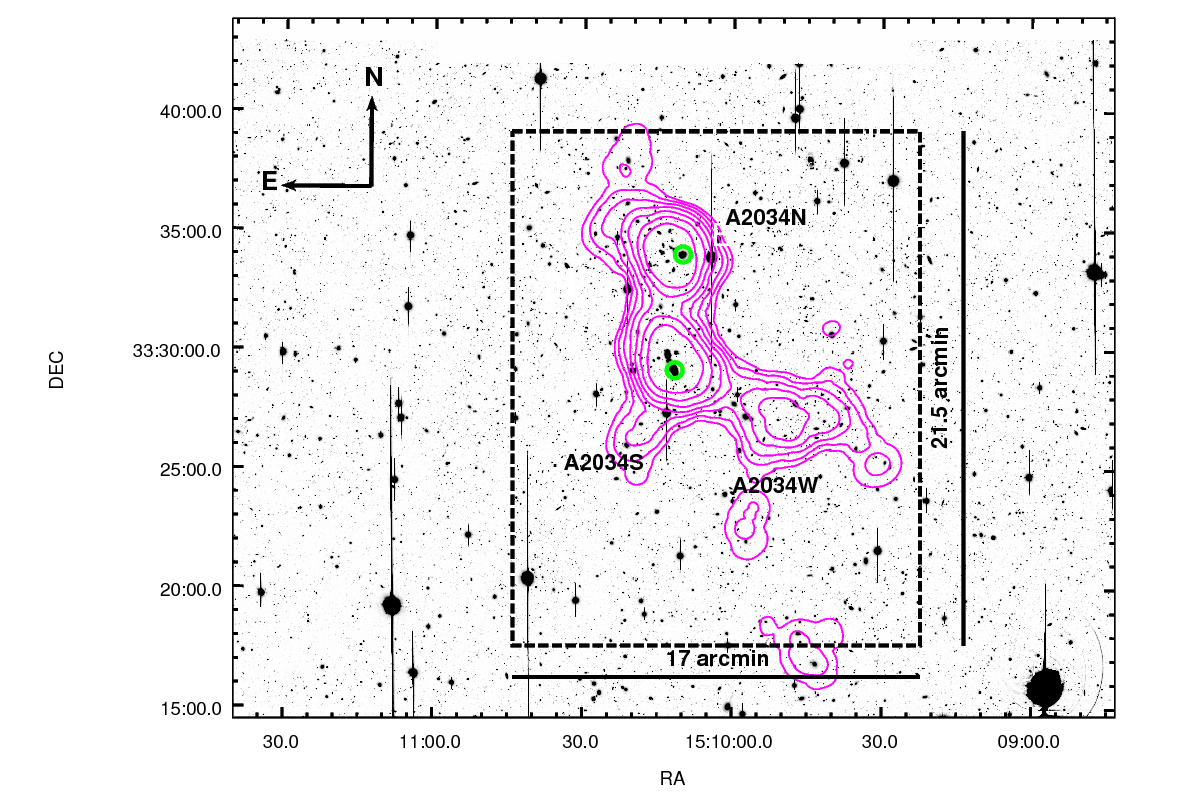} 
\caption{{\it Left:} Numerical density distribution of the 990 red sequence galaxies found through colour-colour statistical subtraction overlaid with the Subaru Suprime-Cam $R_C$ image. Despite the filamentary aspect, we  can see clearly three structures being two of them related to the previous known merging system A2034S and A2034N and a third candidate located at Western which we called A2034W. To carry further analysis on the A2034S\&N dynamics and the pertinence of A2034W to the merging system, we have done a spectroscopic survey on Gemini/GMOS-S whose masks were positioned on the densest regions ({\it blue} boxes). {\it Right: } The numerical density distribution now weighted by the $R_C$ luminosity.  In both maps the contours are logarithm and the BCGs are highlighted with {\it green} circles. For the forthcoming weak lensing analysis, we will focus on the box region ($17 \times 21.5$ arcmin) which encloses the galaxy cluster A2034.}
\label{fig:density}
\end{center}
\end{figure*}

As red cluster members are located preferentially in the inner region \citep{dressler80}, we have identified them by statistical subtraction \citep{Monteiro-Oliveira17b}. We have compared the {\it locus} occupied by galaxies in the $R_C-z'$ versus $B-R_C$ diagram in two different regions: an inner one (dashed  {\it magenta} box in Fig~\ref{fig:A2034.field}) where the red members are supposed to be predominant, and two peripheral regions ({\it blue} dashed box) where the field galaxies will be more numerous. We adopted as a limit for cluster member detection $R_C=22.5$ which is the faintest limit where galaxy counts in the inner region are higher than those in peripheral regions. After the statistical subtraction, we found 990 galaxies within $R_C\leq22.5$. Their number density distribution can be seen in the Fig.~\ref{fig:density}.

The galaxy spatial density distribution shows an elongated filamentary structure related to the merging system A2034S\&N surrounded by a clump at the Western part of the field. We named it as A2034W and its pertinence to the A2034 system will be investigated across this paper. When weighted by $R_C$-luminosity, the numerical density distribution presents a tri-modal behavior. In the merging system, the weighted density is clearly dominated by the BCGs luminosities whereas we do not find a dominant galaxy in A2034W. For further analysis of the mass distribution, we have defined a box enclosing the central part of the image in order to focus our efforts on the aforementioned structures.

\begin{figure}
\begin{center}
\includegraphics[width=1.0\columnwidth, angle=180]{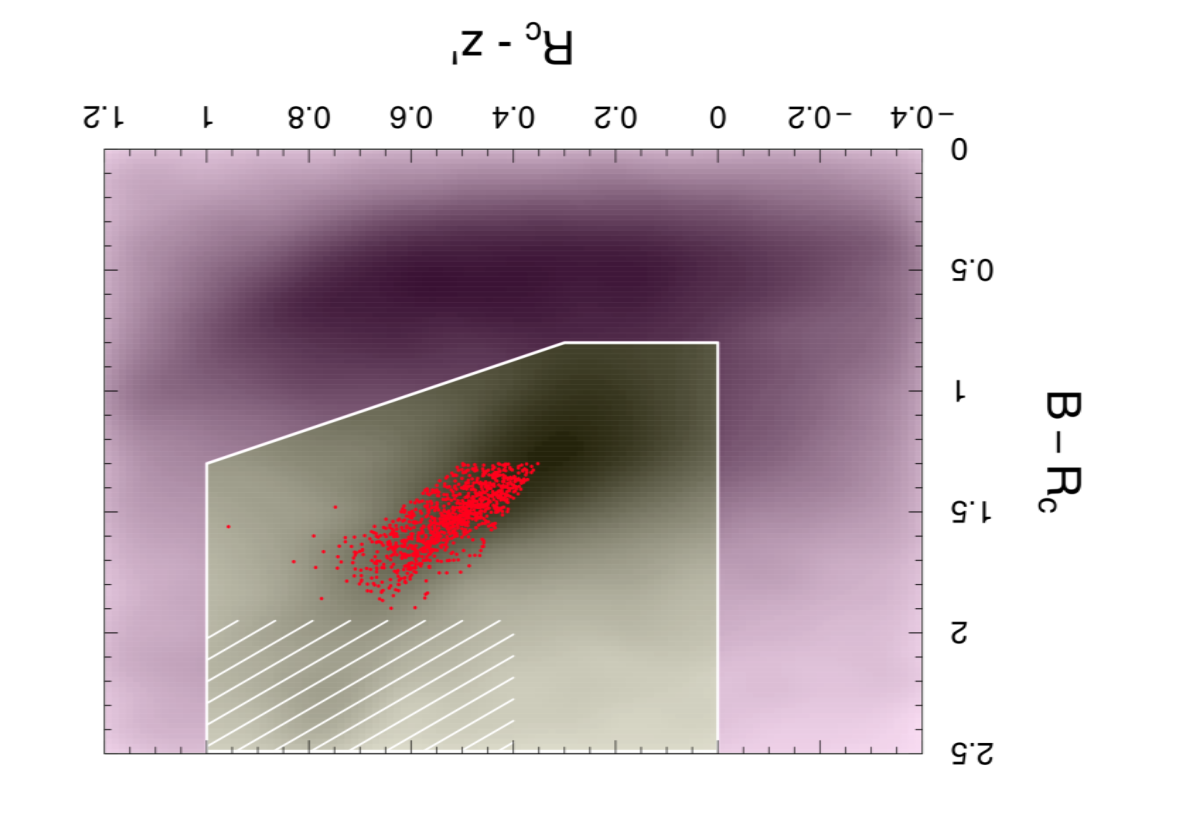}     
\caption{The colour-colour (CC) diagram for the objects classified as galaxies in the A2034 photometric catalogue. The \textit{locus} occupied mostly by the foreground galaxies (central \textit{yellow} region) and the background ones (\textit{magenta}) were obtained by comparison with the work presented by \protect\cite{med10}. The \textit{red} points show the red cluster member galaxies identified after the statistical subtraction. We have also defined a hatched region as the locus of higher redshift early-type galaxies, for the identification of possible background structures \protect\citep{Medezinski18}.}
\label{fig:pop}
\end{center}
\end{figure}

To perform the weak lensing measurements, our selection of the background sample was done taking care to minimize the contamination by foreground and cluster member galaxies. Hence, the background sample was selected satisfying two complementary criteria: ($i.$) galaxies fainter than $R_C=22.5$ and ($ii.$) inside the ``background {\it locus}'' \citep[e.g.][]{capak07, med10} on the $R_C-z'$ versus $B-R_C$ diagram (Fig.~\ref{fig:pop}). Additionally, after the shape measurement process, some of those galaxies will be discarded based on the quality of the obtained ellipticities.

\subsection{Shape measurements}
\label{sec:shape}

Both atmospheric temperature and pressure gradients can change along the light path, which, when arrives at the telescope, suffers an additional diffraction \citep[e.g.][]{schirmer04}. This combined effect of the atmospheric turbulence plus the telescope optics is mathematically described by the point spread function (PSF). In an illustrative way, the PSF acts scattering the photons coming from a point source making it extended following some intrinsic pattern. To perform the unbiased measurement of the galaxies shapes it is necessary to map carefully the PSF and correct for it. This task is done considering the bright unsaturated stars distributed across the field as our target.

\begin{figure}
 \begin{center}
 \includegraphics[width=1.0\columnwidth,angle=180]{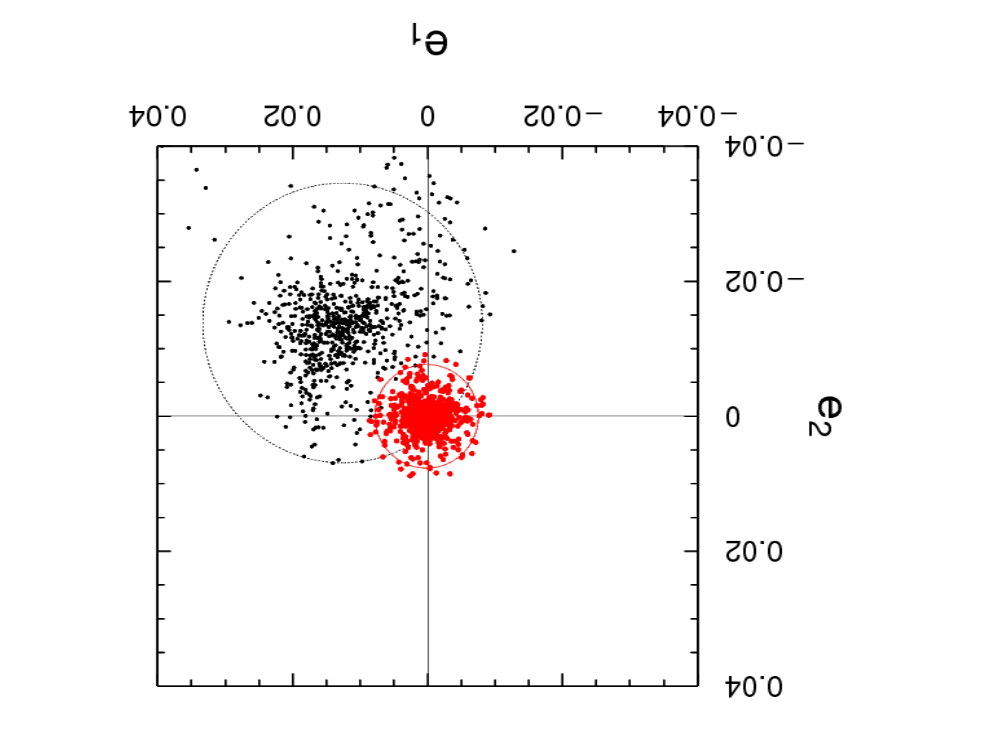}	 
 \caption[]{The {\it black} points show the initial distribution of the ellipticity with $\langle e_1\rangle=0.012$,  $\sigma_{e_1}=0.011$ and  $\langle e_2\rangle=-0.013$, $\sigma_{e_2}=0.012$. For a null PSF, we would found $\langle e_1^{\star} \rangle=\langle e_2^{\star} \rangle=10^{-5}$. The {\it red} points show the residual of the analitical funtion which describes the PSF. We found for $e_1$ and $e_2$  $\langle {\rm res}\rangle=0$ and standard deviation $\sigma_{{\rm res}}=0.003$. Circles represent 95\% of the data.} 
 \label{fig:PSF.model}
 \end{center}
 \end{figure}

The shape measurements for the PSF deconvolution were done by the {\sc im2shape} Bayesian code \citep{im2shape}\footnote{http://www.sarahbridle.net/im2shape/}. It works by modeling each object as a sum of Gaussians with an elliptical basis. Specifically, stars are modeled following single Gaussian profiles and no deconvolution is performed in order to keep the PSF dependent parameters: the ellipticity Cartesian components $e_1$, $e_2$ and the FWHM (full width at half maximum). We have spatially interpolated the discrete set of PSF parameters to generate an analytic function across the entire field. This, in turn, was done in the R environment \citep{R} using the thin plate spline regression \citep{fields}\footnote{We have adjusted $df = 200$ parameters. The procedure has resulted in smoothed PSF parameter surfaces.}. The interpolation was done iteratively, three times, each time removing the 10\% of objects with larger absolute residuals. The final ellipticities measured and the final corresponding residuals after the spatial interpolation are shown in the Fig.~\ref{fig:PSF.model}.

The previously selected background galaxy sample is then obtained by {\sc im2shape} after deconvolving the PSF locally. As an additional quality criterion for our forthcoming lensing analysis, we removed from the sample both galaxies with large ellipticity errors ($\sigma_e>2$) and those with evidence of contamination by nearby objects.

Our final background sample had 26,800 objects, corresponding to a density of 28 gals. arcmin$^{-2}$. The observational parameters are translated into physical quantities through the average critical lensing surface density, $\Sigma_{cr}$. Because we do not have the photometric redshift for each source galaxy, we have estimated its distribution comparing our data with the COSMOS photometric redshift catalog \citep{cosmos}. We selected COSMOS objects using the same colour and magnitude criteria described before to select our background sample. After this, we found $\Sigma_{cr}=4.75(3)\times 10^9$ M$\odot$ kpc$^{-2}$.

\subsection{Mass distribution}
\label{sec:mass.dist}

\begin{figure*}
\begin{center}
\includegraphics[width=1.0\textwidth,angle=180]{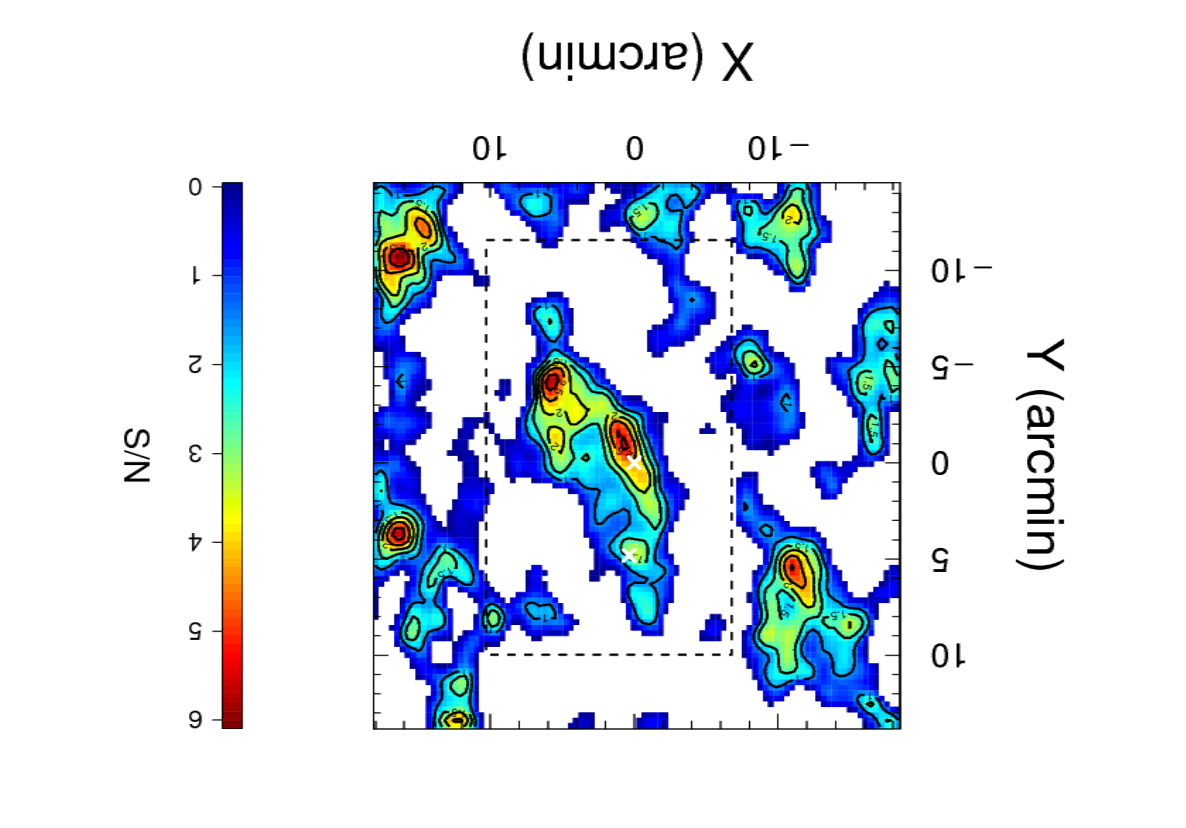}	 
\caption[]{Signal-to-noise ratio (S/N) map obtained through the mass aperture statistic for the entire field covered by the Suprime Cam. Overlaid, are the significance levels in unities of $\sigma_{\rm{S/N}}$ above the mean value $\overline{\rm S/N}$. The central box (dashed {\it black} line) comprises the region previously defined in the Fig.~\ref{fig:density}. The ``$\times$'' are placed at the BCGs position and both are unambiguously related to structures with high S/N. Moreover, the central box region presents other significant peaks whose pertinence or not to the A2034 system will be further investigated through this paper. Several structures with relevant S/N can also be seen near to the field borders. The coordinate origin was set to the position of BCG S.}
\label{fig:SN}
\end{center}
\end{figure*}

As the initial step to investigate the mass distribution, we have mapped the signal-to-noise ratio across the image field starting from the measured ellipticities of the background galaxies. For this purpose, we have referred to the mass aperture statistic \citep{schneider96} following the same procedures adopted in \cite{Monteiro-Oliveira17b} aiming to maximize the detection of dark matter halos at the cluster redshift\footnote{We have adopted an aperture radius of 8 arcmin.}. The final map can be seen in the Fig.~\ref{fig:SN}.

The map shows that the central region, to which we direct all our interest, contains a constellation of peaks surrounding a prominent central one, the position of which nearly matches the BCG S. Besides the central box, the field also presents several peaks with high S/N. However, since we do not find a relevant number of red cluster members outside the box (Fig.~\ref{fig:density}), we speculate that these high S/N regions can be either background massive halos ($\approx 10^{15}$ M$\odot$) or even associations of small ($\lesssim 10^{13}$ M$\odot$) halos \citep{Liu16} aligned. Furthermore, the possibility of them to be artefacts due to image border proximity cannot be ruled out. An attempt to discard the faintest galaxies in order to maximize the S/N in the central region \citep[i.e. to highlight the mass structures candidates to be part of A2034, e.g.][]{Monteiro-Oliveira17b} has proved unfruitful, since the  S/N has remained roughly the same, independently of the adopted cut limit ($R_{C_{{\rm cut}}}\in \{26.5,26,25,25.5,24\}$).

\begin{figure*}
\begin{center}
\includegraphics[width=1.0\textwidth,angle=0]{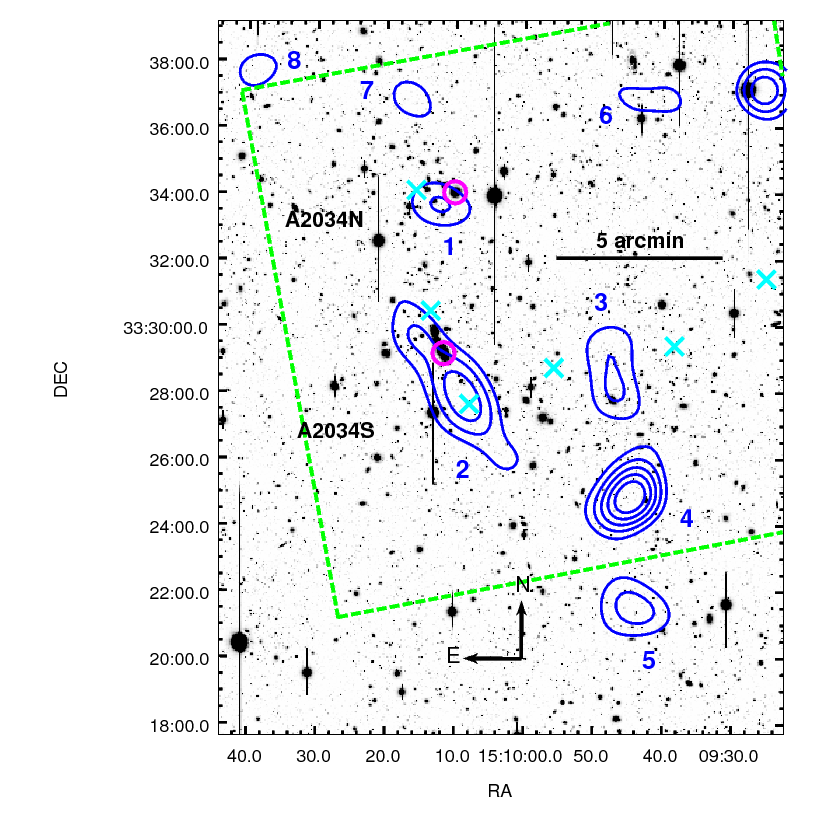}  
\caption{Convergence map obtained by the Bayesian code {\sc LensEnt2} overlaid with our Subaru $R_C$-band image, corresponding to the box region defined in Fig.~\ref{fig:density}. This map was smoothed by a Gaussian intrinsic correlation function (ICF) whose FWHM was set 1.3 arcmin. The mass contours showed in {\it blue} represent the convergence $\kappa$ in unities of $\sigma_\kappa$ starting from the $3\sigma_{\kappa}$ level being $\sigma_\kappa=0.02155$. Numeric labels identify the eight peaks above the $3\sigma_\kappa$ level. The BCGs, identified as {\it magenta} circles, are very close to peaks which we have unambiguously identified as the mass counterparts of the previous structures A2034S and A2034N found in the numerical density map. While the Southern mass concentration appears as an elongated structure, the Northern one appears almost blended with a surrounding clump (withing $2\sigma_\kappa$). Regarding the Western side, three mass concentrations lie in that region and the association with A2034W is not obvious so far. For the sake of comparison, we have included the field where \protect\cite{sevenmergers} have performed a mass reconstruction ({\it green} dashed box). Their most significant peak findings are labelled as ``X''. Overall, their results shows good agreement with those presented here. The slight difference refers to their detection of the clumps \#2 and \#3: whereas they have found a double peak in both, we have found a single one. Notable also, is their non-detection of the clump \#4.}
\label{fig:mass.map} 
\end{center}
\end{figure*}

The next step to proceed is to reconstruct a mass distribution that creates the observed characteristics. For this purpose, we have used the {\sc LensEnt2} Bayesian code \citep{LensEnt2} based on the maximum entropy algorithm \citep{seitz98}. Complementary, the ellipticity field also needs to be smoothed by a filter (Gaussian, in our case) since each galaxy ellipticity is a noisy probe of the shear field. For our data, we have adopted a smooth scale of 1.3 arcmin (80 arcsec), in order to allow both the converge of the {\sc LensEnt2} as well a straightforward comparison with the map produced by \cite{sevenmergers}. A summary of the relevant quantities in the weak lensing mass reconstruction is showed in Tab.~\ref{table:mass.charac} and  the resulting mass distribution is presented in Fig.~\ref{fig:mass.map}.

\begin{table}
\begin{center}
\caption{Characteristics of our weak lensing mass reconstruction: the numerical density of background galaxies available for shape measurements, the FWHM of the smoothing scale and the noise level in the convergence map.}
\begin{tabular}{l c}
\hline
\hline
$N_g$ (gal. arcmin$^{-2}$)  & 28 \\ 
FWHM  (arcsec) & 1.3  \\
$\sigma_\kappa$  & 0.02155\\
\hline
\hline
\end{tabular}
\label{table:mass.charac}
\end{center}
\end{table}

The recovered mass distribution is highly consistent with the S/N map (Fig.~\ref{fig:SN}). The straightforward mass counterparts of A2034S and A2034N, labelled as \#1 and \#2 respectively are noteworthy.  Whereas A2034S appears as an elongated clump in the NE--SW direction, the A2034N mass counterpart seems to be blended with a neighbouring mass concentration.

In both maps there are several mass clumps comparable to the cluster ones with no clear optical counterparts (excess of galaxies with the same colour as the cluster ellipticals). 
It can be understood partially given the relatively low redshift of A2034 ($\bar{z}=0.114$). Even moderate structures at $0.2<z<0.4$ will be as effective as lenses as A2034 itself. What remains to be discussed is whether those peaks correspond to actual structures, bound or  projected along the line-of-sight, or fake peaks due to the noise in the data \citep{Wei18}.

A remarkable feature on those maps are the three prominent Western clumps (\#3, \#4 and \#5).  An overall comparison with the mass maps shown by \cite{sevenmergers}, (see the {\it green} dashed box in Fig.~\ref{fig:mass.map}) shows qualitative agreement. There are two slight differences, however: ($i.$) the detection of two (instead one) structures regarding our peaks \#2 (their ``C''and ``S'') and \#3 (their ``W1'' and ``W2'') and ($ii.$) the absence of the peak \#4 in \cite{sevenmergers}.

\begin{figure}
 \begin{center}
 \includegraphics[width=1.0\columnwidth,angle=180]{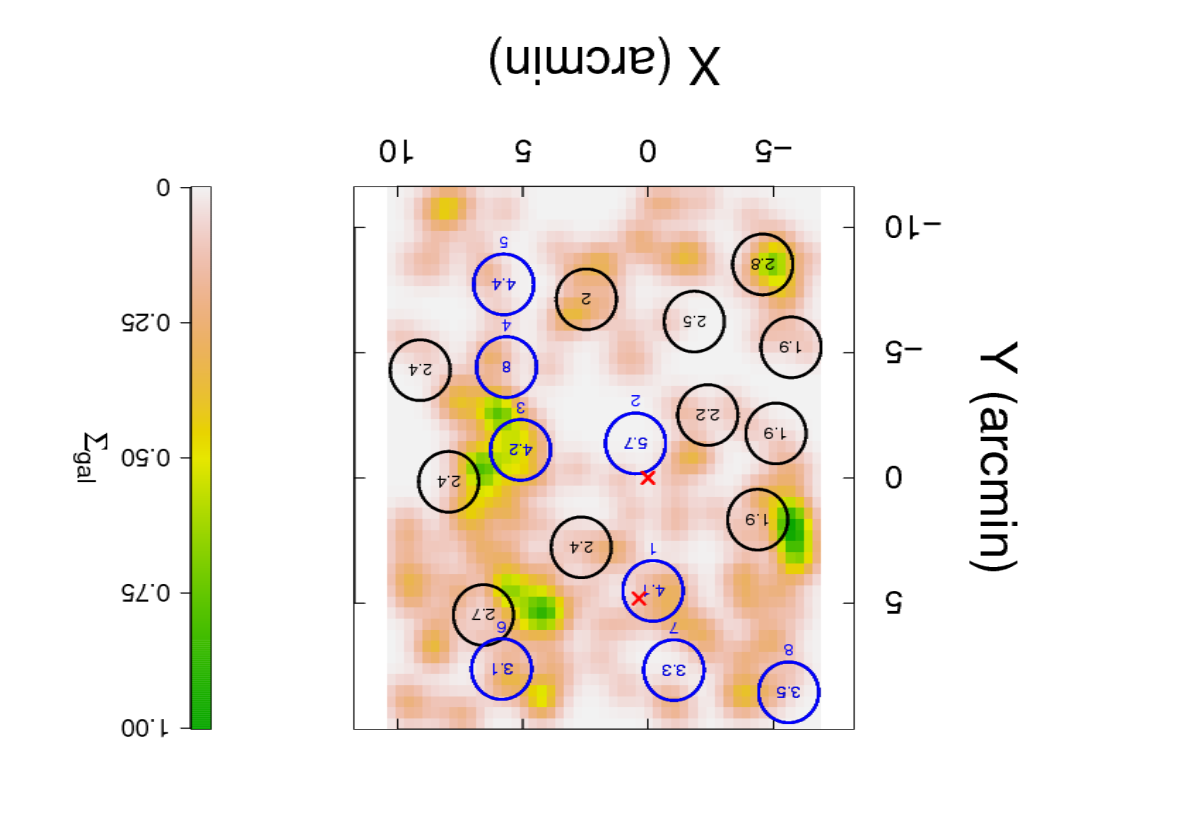}	 
 \caption[]{Mass clumps identified by our peak finder (circles) and their respective significance (internal labels). In {\it blue} are highlighted the most significant peaks of the field ($\nu>3$) labelled as \#1:8. The position of the BCGs are marked as a '$\times$'. The background map represents the projected spatial distribution $\Sigma_{\rm gal}$ (in relative unities) of the redder galaxies located at the same branch of the cluster members in the CC diagram (Fig.~\ref{fig:pop}). We found most of the relevant peaks (excluding those unambiguously regarding A2034N and A2034S) not related to obvious optical counterparts. An exception is the peak \#3, surrounded by an excess of galaxies.}
 \label{fig:optical.counter}
 \end{center}
 \end{figure}

Disregarding their nature, we have carried out a quantitative characterization of the mass clumps. Our self-made peak finder searched for local maxima in the convergence map using a moving circular window with a radius of $\sim1.2$ arcmin. The peak centre was then calculated based on the pixel-weighted mean inside the window and the significance $\nu$ was defined as the ratio between the central convergence $\kappa_{\rm max}$ and the noise level $\sigma_\kappa$. The mass clumps identified can be seen in the Fig.~\ref{fig:optical.counter}. A cut at $\nu>4$ is enough to select the A2034 related structures in spite of  the real structures  (those not related to the noise) can arise with $\nu>3$ \citep[e.g.][]{Gavazzi07, Wei18}.

Aiming to explore the nature of the ``orphan'' mass clumps, we first tested whether those can be associated with background galaxy clusters.
For this, we have considered the galaxies belonging to a redder locus in the CC diagram (hatched region in Fig.~\ref{fig:pop}) where higher redshift red-sequence galaxies are expected to lie \citep{Medezinski18}. The spatial projected density of those galaxies is showed in Fig.~\ref{fig:optical.counter}. As we can see, most of the mass peaks are not straightly related to any galaxy concentration. An exception is the peak \#3 which appears to be surrounded by an excess of galaxies.

We then verified the probability of occurrence of fake peaks generated by the noise in the convergence map. We have addressed this issue by generating a sample of galaxies at the same position of the original catalogue but with their ellipticities rotated by a random angle each. This procedure is intended to erase the cluster signal, allowing to quantify the noise level and detect possible fake peaks generated by noise fluctuations.  We then performed 100 realizations of the mass reconstruction and peak finding which, following  e.g. \cite{Martinet16} is enough to provide reliable results.  We show the results of this analysis in Tab.~\ref{table:peak.detection}.  As it can be seen there, most of the detected peaks above $3\sigma_\kappa$ must be real as the expected number of fake peaks is below 1 at 3$\sigma$.

\begin{table}
\begin{center}
\caption{Number of peaks detected in function of their statistical significance. We also show the expected number of fake peaks obtained through a resampling of the galaxy ellipticities.}
\begin{tabular}{l c c }
\hline
\hline
Threshold& Original detection & Expected fake peaks \\
\hline
$>1\sigma_\kappa$  & 19 & $13.5\pm5.4$ \\ 
$>2\sigma_\kappa$  & 15 & $1.6\pm2.1$  \\
$>3\sigma_\kappa$  & 8  & $0.1\pm0.3$ \\
$>4\sigma_\kappa$  & 5  & $0.02\pm0.14$ \\
\hline
\hline
\end{tabular}
\label{table:peak.detection}
\end{center}
\end{table}

\begin{figure}
 \begin{center}
 \includegraphics[width=1.0\columnwidth,angle=0]{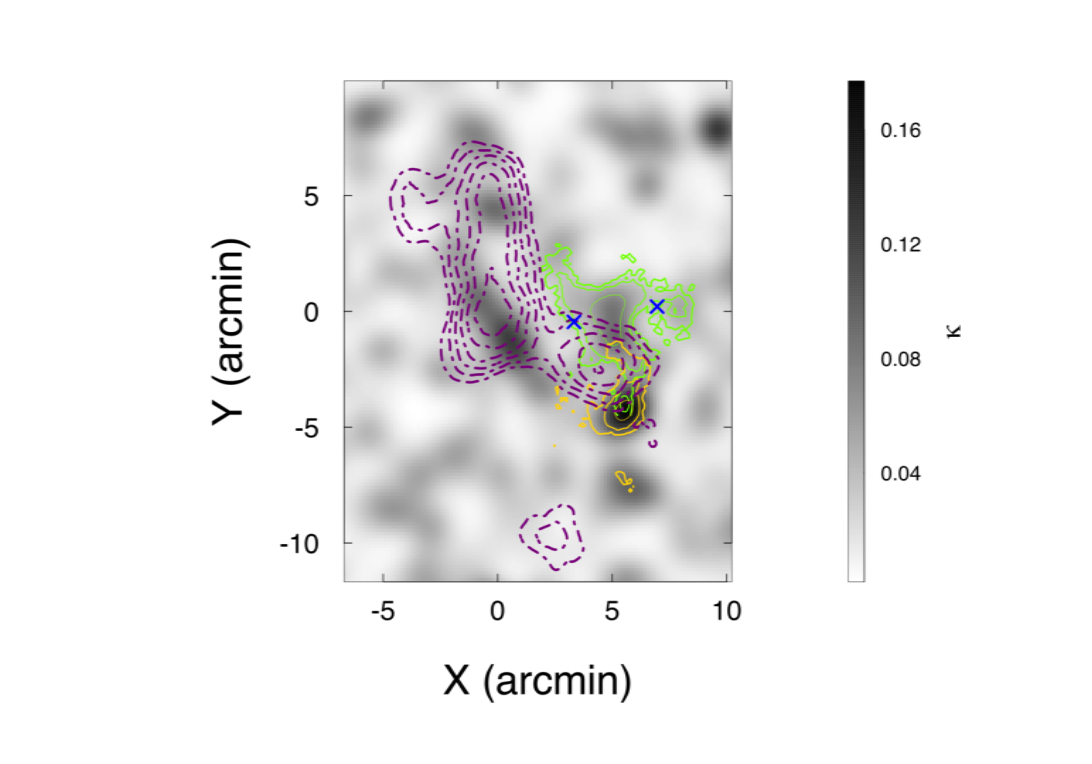}	 
 \caption[]{Original convergence map ({\it gray} scale) overlaid by the results of the $10^4$-bootstrap resamplings showing the 68\%, 95\% and 99\% c.l. of the centre position of the peaks \#3 (green) and \#4 (yellow). For a comparison we also show the numerical density distribution of the red sequence galaxies presented in Fig.~\ref{fig:density}. Note that the Western galaxy clump A2034W is located preferentially between the two mass clumps. The ``$\times$'' corresponds to the position of the Western mass clumps found by \cite{sevenmergers}.}
 \label{fig:boot.peaks}
 \end{center}
 \end{figure}

Having established the reality of the mass clumps, we turn our attention to investigate the mass counterpart of A2034W. In order to check the uncertainties in the position of the detected peaks, we have performed a bootstrap resampling, generating $10^4$ shear fields (keeping the same ellipticity and position but allowing for repetition of data) and computing the correspondent mass maps. For each realization we have searched for the closest peak position relative to those previously identified in the original map (Tab.~\ref{A2034.tab:masses}). The results are presented in Fig.~\ref{fig:boot.peaks}.

The Western region of the galaxy numerical density distribution, also presented in the Fig.~\ref{fig:boot.peaks}, is located midway the peaks \#3 and \#4 making dubious a direct association to each one. In fact, the mass reconstruction provided by  \cite{sevenmergers} is also unable to recover the mass counterpart of A2034W. As we can see in Fig.~\ref{fig:boot.peaks}, the position of their subclumps W1 (Eastern) and W2  (Western) are compatible with our clump \#4 within 99\% c.l. However, both are displaced relative to the A2034W centre.

We will resume this discussion later in this work. For now, we  concentrate our efforts to investigate whether A2034 can be described as a bimodal merger between the subclusters A2034S and A2034N as suggested by previous studies accounting for the dynamical state of the ICM \citep[e.g.][]{owers14}.

\subsection{Mass distribution modelling}
\label{sec:mass.model}

Following our proposal of providing a comprehensive description of the mass distribution in A2034, we have performed the mass modelling. For this, we have considered strictly for the data comprised inside the box region defined in the Fig.~\ref{fig:density} \citep[e.g.][]{Monteiro-Oliveira17b}.

The mass distribution model takes into account  the peaks with $\nu>3$, each one simultaneously fitted by eight universal NFW profiles \citep[uNFW;][]{nfw96,nfw97}. Complementary, we have employed a  truncated NFW profile  \citep[trNFW;][]{Baltz09}. This variant is intended to be a better description of merging halos instead the uNFW \citep[e.g.][]{Medezinski16}. 

Due to the fact that we are dealing with a multiple system, where spherical symmetry is absent, it is easier to work with the Cartesian components of the shear, $g_1$ and $g_2$, instead of the usual tangential component $g_+$. The transformation is done by multiplying $g_+$ by the lensing convolution kernel,
\begin{equation}
D_1 = \frac{y^2 - x^2}{x^2 + y^2}\mbox{,}
\quad
D_2 = \frac{-2xy}{x^2 + y^2},\\
\label{A1758.eq.kernel}
\end{equation} 
where $x$ and $y$ are the Cartesian coordinates relative to the respective lens centre.

For each background galaxy, the effective shear is then written as a sum of each mass clump,
\begin{equation}
g_i = \sum_{k=1}^{N_{\rm clumps}}  g_i^{k},
\label{eq:g_sum_clumps}
\end{equation}
with $i\in\{1,2\}$. We can now write the  $\chi^2$-statistics as
\begin{equation}
\chi^2=\sum_{j=1}^{N_{{\rm sources}}} \sum_{i=1}^{2}  \frac{(g_i-e_{i,j})^ 2}{\sigma_{int}^2+\sigma_{ obs_{i,j}}^2}, \label{eq:chi2_d}
\end{equation}
where $g_i$ is the effective shear (Eq.~\ref{eq:g_sum_clumps}), $e_{i,j}$ is the measured ellipticity given by {\sc im2shape}, $\sigma_{obs_{i,j}}$ ellipticity uncertainty and $\sigma_{int}$ is the uncertainty related to the intrinsic galaxy shapes, estimated as $\sim 0.35$ for our data. The likelihood can be written as,
\begin{equation}
\ln \mathcal{L} \propto - \frac{\chi^2}{2}\mbox{.} 
\label{eq:likelihood.d}
\end{equation}
%

Aiming to reduce the number of degrees of freedom in our model, we have assigned the NFW-concentration parameter $c$ using the \cite{duffy08} $M_{200}-c$ relation,
\begin{equation}
c=5.71\left(\frac{M_{200}}{2\times10^{12}h^{-1}M_{\odot}}\right)^{-0.084}(1+z)^{-0.47},
\label{eq:duffy_rel}
\end{equation}
and we have considered all mass clumps located at the same A2034 redshift ($z\approx0.114$). Each clump centre was fixed in the coordinates determined by our peak finder algorithm. Additionally, to accelerate the model convergence, we applied an uniform prior $\mathcal{P}(M_{200})$, $0<M_{200}\leq 8\times 10^{15}$ $M_{\odot}$. Regarding the trNFW model, we have adopted two cases: a cut at  $R_{200}$ \citep[hereafter trNFW; e.g.][]{dawson} and a truncation at $2R_{200}$ \citep[hereafter 2trNFW; e.g.][]{Oguri11}.

Finally, the posterior for the model can be written as
\begin{equation}
 {\rm Pr}( M_{200}|\rm data) \propto  \mathcal{L}({\rm data}|M_{200})\times \mathcal{P}(M_{200})\mbox{.}
\label{eq:posterior}
\end{equation} 

\subsection{Results}
\label{sec:mass.results}

The posterior given by Eq.~\ref{eq:posterior} was probed by the MCMC (Monte Carlo Markov chain) algorithm with a simple Metropolis sampler implemented in the {\sc R} package {\sc MCMCmetrop1R} \citep{MCMCpack}. The method generated four chains with $1\times10^5$ iterations each allowing an additional $1\times10^4$ iterations as ``burn-in'' to ensure the chains fully represent the stationary state. To check the convergence of the final chain, we computed the potential scale factor $R$, as implemented in the {\sc Coda} package \citep{coda}. It has shown that the final combined chain is, within 68\% c.l, convergent ($R\approx1.0$).

In Tab.~\ref{A2034.tab:masses} we present the measured masses marginalised over all other parameters, being the median of the distribution considered as the fiducial value. Noticeable is the consistency among the results presented by the different NFW models. To quantify this assessment we have computed two widely used statistical indexes for measuring the relative model quality, the Akaike information criterion (AIC) and the Bayesian information criterion (BIC). Concerning each criterion, the lowest index value presented among the different models points in favour of the preferred one. For the trNFW model we found $\Delta{\rm BIC}= \Delta{\rm AIC}=2$ in relation to the uNFW, whereas the indexes are not able to disentangle the best model when we have compared the  uNFW and the 2trNFW ($\Delta{\rm BIC}= \Delta{\rm AIC}=0$). Consequently, we have adopted the universal single NFW as our fiducial model hereafter.

\begin{table*}
\caption[]{Summary of the statistically relevant mass clumps. The first five columns refer to the clump ID [1], the peak centre coordinates [2-3], the peak significance $\nu=\kappa_{\rm max}/\sigma_\kappa$ [4] and the mean convergence inside a circular region [5]. The last three columns corresponds to the masses obtained through our different models, respectively a universal NFW (uNFW) [6], a truncated NFW at $R_{200}$ (trNFW) [7] and at $2R_{200}$ (2trNFW) [8]. The fiducial values correspond to the median of the MCMC chains. The error bars correspond to the  68\% c.l. Please note that, apart from A2034S and A2034N, this masses will only correspond to the ``true masses'' if the clump is located at the same cluster redshift.}
\label{A2034.tab:masses}
\begin{center}
\begin{tabular}{c c c c c c c c }
\hline
\hline 
 Clump & $\alpha$ (J2000) & $\delta$ (J2000) & $\nu$ & $\bar{\kappa}$ ($<{\rm 1.2}$ arcmin) & $M_{200} [\rm uNFW]$ ($10^{14}$ M$\odot$) & $M_{200}  [\rm trNFW]$ ($10^{14}$ M$\odot$) & $M_{200}  [\rm 2trNFW]$ ($10^{14}$ M$\odot$)\\
\hline
	1$^\star$ &  15:10:13 & +33:33:43 & 4.1 & 0.055 & $1.08_{-0.71}^{+0.51}$ & $1.39_{-0.85}^{+0.59}$  & $1.16_{-0.74}^{+0.54}$ \\[5pt] 
	2$^\diamondsuit$ &  15:10:09 & +33:27:50 & 5.7 & 0.085 & $2.35_{-0.99}^{+0.84}$ & $2.80_{-1.10}^{+0.94}$   & $2.47_{-1.03}^{+0.88}$\\[5pt]	
	3 &  15:09:47 & +33:28:07 & 4.2 & 0.064 & $1.63_{-0.94}^{+0.68}$ & $1.97_{-1.08}^{+0.78}$   & $1.73_{-1.02}^{+0.69}$ \\[5pt]
	4 &  15:09:45 & +33:24:49 & 8.0 & 0.095 & $1.59_{-1.03}^{+0.66}$ & $1.71_{-1.09}^{+0.71}$   & $1.62_{-1.03}^{+0.70}$ \\[5pt]	
	5 &  15:09:44 & +33:21:31 & 4.4 & 0.064 & $1.78_{-1.13}^{+0.79}$ & $1.83_{-1.17}^{+0.81}$   & $1.77_{-1.17}^{+0.77}$	\\[5pt]	
	6 &  15:09:44 & +33:36:52 & 3.1 & 0.051 & $0.55_{-0.47}^{+0.28}$ & $0.68_{-0.55}^{+0.33}$   & $0.59_{-0.49}^{+0.30}$ \\[5pt]
	7 &  15:10:17 & +33:36:52 & 3.3 & 0.049 & $0.72_{-0.61}^{+0.35}$ & $0.84_{-0.68}^{+0.39}$   & $0.76_{-0.64}^{+0.36}$ \\[5pt]	
	8 &  15:10:39 & +33:37:44 & 3.5 & 0.043 & $5.74_{-2.39}^{+1.88}$ & $6.09_{-2.54}^{+2.04}$   & $5.84_{-2.47}^{+1.98}$ \\	
\hline
\hline
\multicolumn{8}{c}{$^\star$ A2034N}\\
\multicolumn{8}{c}{$^\diamondsuit$ A2034S}\\
\end{tabular}
\end{center}
\end{table*}

We found $M_{200}^S=2.35_{-0.99}^{+0.84}\times 10^{14}$ M$_{\odot}$ and $M_{200}^N=1.08_{-0.71}^{+0.51}\times 10^{14}$~M$_{\odot}$ leading to a total mass for A2034S\&N M$_{200}^{\rm S+N}=3.54_{-1.17}^{+0.95}\times10^{14}$~M$_\odot$. Our posterior also show that the subcluster  A2034S is more massive than A2034N in  $85\%$ of the MCMC realisations. The total mass estimated is coherent with the value proposed by \cite{delliou15}, also based on weak lensing measurements (according to the Tab.~\ref{tab:mass.comp}). However, our total mass is smaller than those estimated by \cite{geller13} and \cite{owers14} using the caustic methodology. This discrepancy can arise from the misidentification of the cluster centre, which is an important parameter required by the caustic technique and is not obvious in the case of merging systems. We also warn that the presented total mass does not take into account the counterpart of A2034W which tends to diminish the appointed discrepancy.

\begin{figure*}
\begin{center}
\includegraphics[angle=180, width=1.1\textwidth]{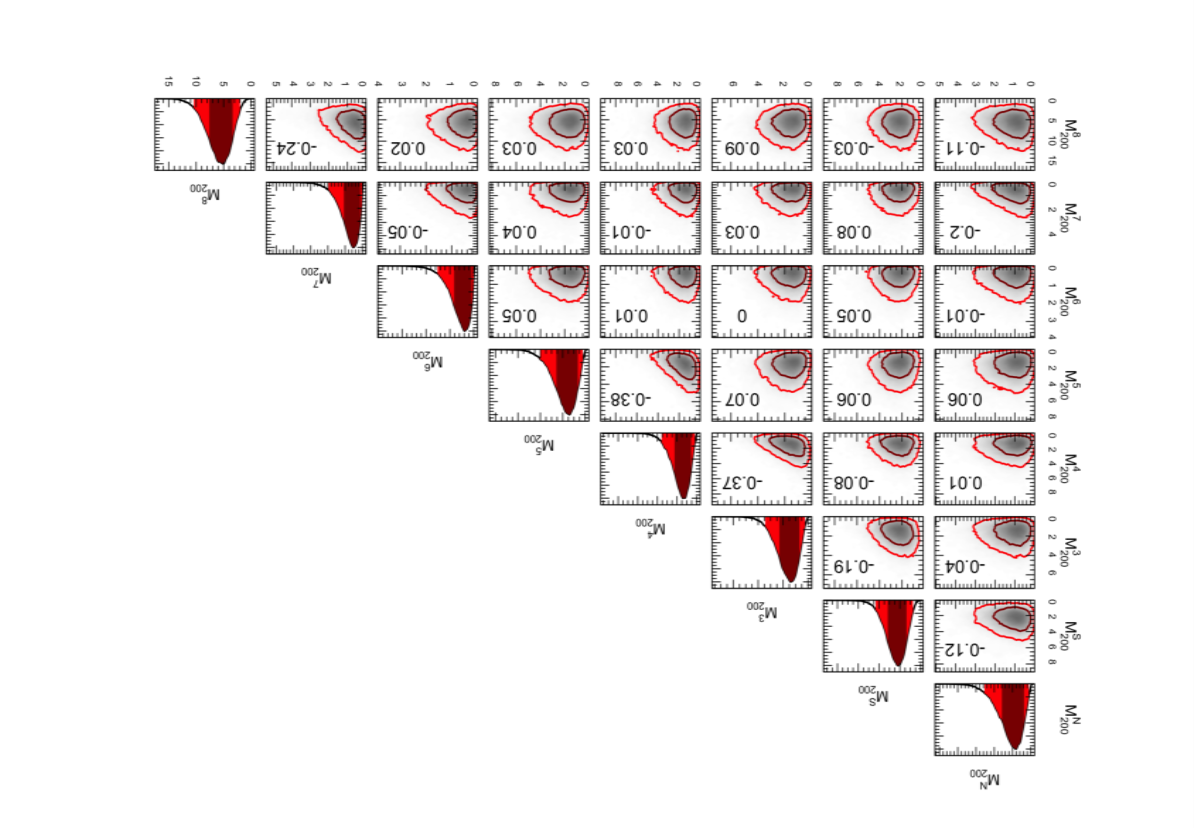}
\caption{Posterior of the 8-NFW modelled masses in our Bayesian model. In the diagonal we present the marginalised posterior for the individual mass parameters. The number shows the Pearson's correlation for each pair.} 
\label{fig:mass.posterior}
\end{center}
\end{figure*}

The marginalised posteriors are shown in Fig.~\ref{fig:mass.posterior} as well as the Pearson correlation coefficient among them. In general the (anti) correlation is negligible, except between the neighbour pairs A2034N \& \#2 ($-0.2$) and A2034S \& \#3 ($-0.19$), the Western clumps \#3 \& \#4 ($-0.37$) and \#4 \& \#5 ($-0.38$) and the Northern clumps \#7 \& \#8 ($-0.24$). 

An alternative uNFW model taking into account only the five peaks with $\nu>4$ was also computed. The resultant masses were fully consistent with the values presented in Tab.~\ref{A2034.tab:masses} except for a noticeable difference: the mass attributed to A2034N clump is now twice its value in the 8-peaks model.  This behaviour is easily explained by the existence of a moderate anti-correlation between the clump associated with A2034N and its neighbours, the clump \#7 (Fig.~\ref{fig:mass.posterior}). Statistically, whereas the BIC criterion does not distinguish between the models involving different number of parameters  ($\Delta{\rm BIC}\approx 0$), the AIC criterion points straightforward in favour of the 8-peaks model, with $\Delta{\rm AIC} = 26$.

We can now turn our attention to compare the relative position of both total mass and the ICM distributions in the region of the merging system A2034S and A2034N. Regarding the hot X-ray emitting gas distribution, we have downloaded 4 pointings publicly available made with the X-ray telescope \textit{Chandra}: IDs 2204, 12885, 12886, and 13192 (PIs. G. Sarazin and P. Nulsen). Periods with high particle background (flares), which are caused by protons accelerated by the Sun, were excluded from the event files using the lc\_clean script from the Chandra X-ray Center (CXC). The merged individual observations have an effective total exposure time of 243~ks. All these exposures were done in a very faint mode with ACIS-I. We have produced an exposure-map corrected (flat) image in broad band (0.5--7.0~keV) with the script \texttt{merge\_obs} available from the CXC, with a pixel scale of $0.984$ arcsec. We also generated an adaptive smoothed image in the same energy band with the tool \texttt{dmimgadapt} from CIAO 4.8.

\begin{figure}
\begin{center}
\includegraphics[angle=0, width=\columnwidth]{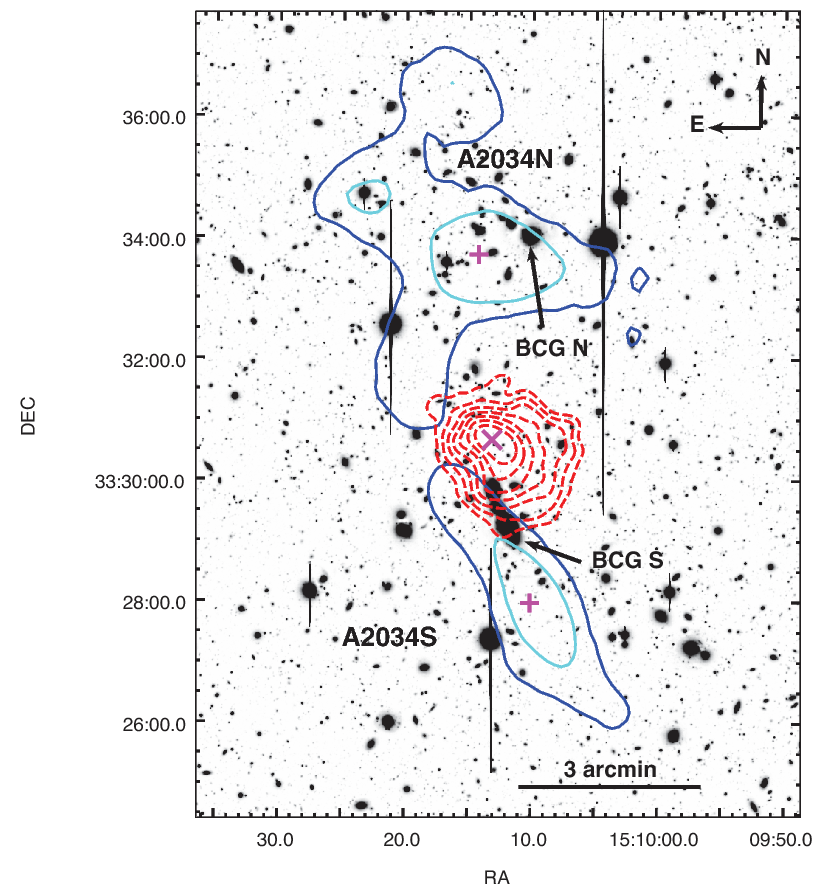}
\caption{ $R_C$ image overlaid with the brightest region of the X-ray emission mapped by {\it Chandra}  ({\it red} contours). In {\it cyan} and {\it blue} are presented the confidence levels (respectively 1$\sigma$ and 2$\sigma$) of the mass peaks related to A2034S and A2034N obtained via bootstrapping with $1\times10^4$ resamplings. The exact position of the original mass peaks are marked as ``+'' whereas the single X-ray peak appears as an ``X''. The position of both BCGs are consistent with their related mass peaks being 1$\sigma$ for the BCG N and $2\sigma$ for the BCG S. On the other side, the X-ray peak is detached $91\pm1$ from the BCG S and $169^{+48}_{-42}$ arcsec from the Southern mass peak. The present image corresponds to a zoomed region of the Fig.~\ref{fig:mass.map}.} 
\label{fig:mass.peak.uncer}
\end{center}
\end{figure}

Due to the complexity of the field, we were not able to set the mass centre positions as free parameters in our model \citep[e.g.][]{Monteiro-Oliveira17a}. As an alternative to estimate the mass centre uncertainties, we have done $10^4$ bootstrap re-samplings of the ellipticity field. For each one, we have obtained the respective mass distribution (as described in Sec.~\ref{sec:mass.dist}) and mapped the position of the nearest peak in relation to those previously identified (Fig.~\ref{fig:mass.map}). We present the confidence contours and the comparison with the ICM X-ray emission in Fig.~\ref{fig:mass.peak.uncer}.

Both BCGs, which are 4.9 arcmin apart from each other, have their location coincident with their related mass peaks. For A2034N, this agreement is within 1$\sigma$ whereas in A2034S it happens within $2\sigma$ being the BCG/mass peak separation equal to $84^{+37}_{-32}$ arcsec. Given the relatively low mass of A2034N and the proximity with a neighbouring structure (\#7), the determination of its peak position is less precise than we found for A2034S. In some cases, the mass peak of A2034N becomes confused with its Northern neighbor which is far $199_{+45}^{-41}$ arcsec from the BCG N. In relation to the mass peak, the BCG N is found $57^{+36}_{-34}$ arcsec away from it.

The single X-ray peak presents a clear offset from the other cluster components appearing detached $91\pm1$ arcsec from the BGC S and $169^{+48}_{-42}$ arcsec from the mass peak related to A2034S. Regarding A2034N, previous studies \citep[][]{kempner03,owers14} have proposed that the gas content of this subcluster was stripped out forming currently a shock front \citep{owers14} behind the BCG N. Therefore, the merger between A2034S and A2034N can be classified as a doubly dissociative system resembling the remarkable Bullet Cluster \citep[e.g.][]{clowe04,clowe06}.

\section{Dynamical analysis}
\label{sec:dynamical}

Following our goal to characterize the current state of the A2034S\&N merging system we will check if the dynamics of the galaxy members, as revealed by their radial velocities, match the previous mass structures.

\subsection{Spectroscopic data}
\label{sec:spect.data}

To address the A2034 dynamics, we observed the target during 3.5 h in 2013A (GN-2013A-Q-36, PI: Rog\'erio Monteiro-Oliveira) with the Gemini Multi-Object Spectrograph mounted at the Gemini North telescope. We used three slit masks being two targeting the merging system A2034S and A2034N and a third positioned in the Western region where we and \cite{sevenmergers} found some mass concentration (see the slit masks positions in the Fig~\ref{fig:density}). We totalled  119 galaxies observed, most of them selected as a red sequence member (Fig.~\ref{fig:pop}). We also targeted some bluer galaxies in order to maximize the number of objects per mask whose total integration time was 70 minutes each. We used the R400 grating and 1.0 arcsec wide slits, which lead to spectra with $\Delta\lambda\approx8$\AA~ in $6500$ \AA. Data standard reduction and calibration in wavelength was performed with {\sc gemini.gmos} {\sc IRAF} package.

For galaxies with absorption lines, the radial velocities were obtained through the cross-correlation technique \citep{td79} implemented in the task {\sc xcsao} from {\sc RVSAO} package \citep{rvsao}. The observed spectra were correlated with several galaxy spectra templates of similar resolution but large S/N. At the end, the code output a redshift for each template. We selected among them taking as a primary criterion the cross-correlation coefficient $R\ge3$ \citep{td79}. Complementary, we have done a visual inspection of the spectra superimposed with the positions of the main spectra lines. This step is important to reject some wrong solutions found occasionally in low S/N spectra. At the end, we found 91 measured redshifts.

The field of A2034 was previously observed and a subsequent search in the NASA Extragalactic Database (NED)  revealed 95 galaxies with spectroscopic redshifts. The four galaxies in common with our {\sc  gemini} catalogue showed a redshift residual comparable to zero. Our final catalogue has 182 galaxies with spectroscopic redshifts (see Appendix~\ref{ap:catalogue}).

\subsection{Searching for substructures}
\label{sec:search.sub}

\begin{figure}
\begin{center}
\includegraphics[width=1.0\columnwidth, angle=180]{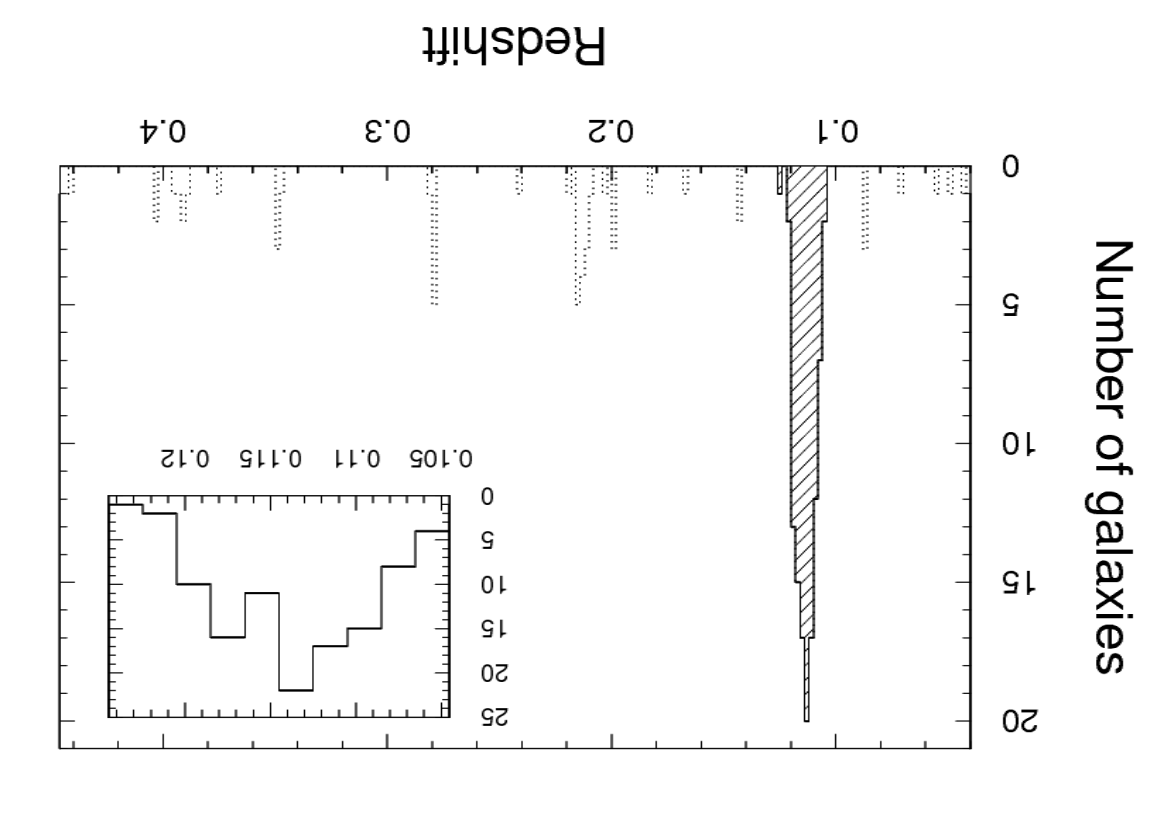}     
\caption{Redshift distribution in A2034 field. To make this plot clear, we removed ten galaxies with $z>0.45$. The spectroscopic cluster members were selected after a $3\sigma$-clipping procedure. They are highlighted  as dashed lines and are shown in more detail in the upper right panel. The 106 selected members have $\bar{z}=0.1135$ and $\sigma/(1+\bar{z})=1086$ km s$^{-1}$.}
\label{fig:redshifts}
\end{center}
\end{figure}

After implementing a 3$\sigma$-clipping cut \citep{3sigmaclip}, we found 106 spectroscopic members whose redshift distribution is shown in Fig.~\ref{fig:redshifts}.  This sample has  $\bar{z}=0.1135$ with $\sigma/(1+\bar{z})=1086$ km s$^{-1}$ and is, within 99\% c.l., Gaussian distributed. The $\Delta$-test \citep{ds} points to the absence of substructures with 99\% c.l. \citep[p-value~=~0.16;][]{hou12}.

The absence of any substructure in the redshift space is in contrast with the observed scenario both in X-rays and mass distribution. There are two main explanations for that: first, we could be observing the colliding system close to their apoapsis, when the relative velocity becomes null \citep[e.g.][]{golovich16} and/or the merging is taking place near to the plane of the sky \citep[e.g.][]{Monteiro-Oliveira17a,Monteiro-Oliveira17b}. Given the presence of X-ray features like a shock front behind A2034N \citep{owers14} we believe that the second proposition is more plausible, otherwise the merging imprints on the ICM would not be seen \citep[e.g.][]{Molnar18}.

To identify how the galaxy distribution behave in comparison to the other components (ICM and total mass), we will turn to the  spatial projected red sequence galaxy distribution. Our sample was composed of the spectroscopically confirmed members plus all objects from the cluster red sequence up to $R_C\leq22.5$ inside the box region defined in Fig.~\ref{fig:density}. As pointed out, the galaxy field is very irregular, showing a filamentary structure with three densest regions surrounded by few others.

\begin{figure*}
\begin{center}
\includegraphics[width=1.0\textwidth,angle=180]{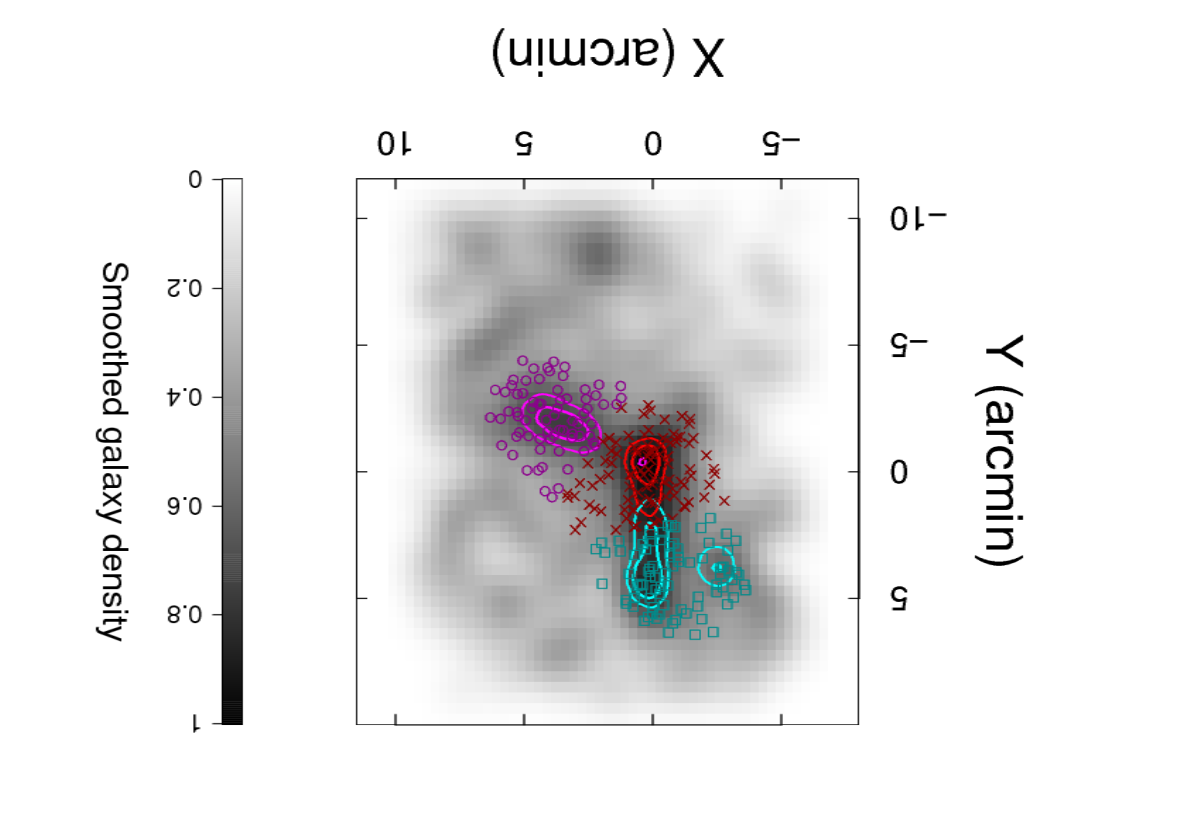} 	 
\caption{Smoothed map of the red sequence members spatial distribution obtained by the {\sc Bkde2d} implemented on the {\sc R} environment. This linear map shows, beyond the remarkable trimodal concentration, a few other clumps surrounding the previous ones. The most prominent galaxy concentrations were quantitatively identified at 3.6$\sigma$, 3$\sigma$ and 2.5$\sigma$ above the mean respectively for the previous clumps named A2034S, A2034N and A2034W. The contours are showing the uncertainty (1--2$\sigma$) of each peak position after $10^5$ resamplings of the original data after a small random change within a scale of 50 arcsec. The points correspond to the galaxies belonging to the densest part corresponding to members of A2034S (diamond), A2034N (``X'' ) and A2034W (``+'') which were classified by the $2$-dimensional Gaussian mixture model algorithm {\sc Mclust}.}  
\label{fig:peak.significance}
\end{center}
\end{figure*}

In order to identify the most significant concentrations in the spatial projected galaxy distribution, we have generated a smoothed field through the {\sc R} function {\sc bkde2d}. It uses a bivariate Gaussian kernel having, in this case, 50 arcsec (103 kpc, for both $x$ and $y$ directions). Afterwards, our peak finder searched for the local maximum inside a moving window equally sized. At the end we found that the merging components A2034S and A2034N were detected respectively $3.6\sigma$ and $3.0\sigma$ above the mean density. The third most significant concentration is that previously named A2034W, being detected at $2.5\sigma$. Regarding the other density peaks, they show significance not larger than $1.8\sigma$. To certify that the previous peaks were not detected by chance,  we have generated $1\times10^5$ resamplings of the original galaxy distribution each one a little bit different due to a small random perturbation generated through the {\sc R} function {\sc Rjitter} within a scale of 50 arcsec\footnote{This is a sensible parameter. If small, we are not resampling the cluster. On the other side, large values can disfigure the cluster appearance. After some test, we found 50 arcsec radius as the best compromise ensuring that the spatial correlation among the objects is preserved at the same time the cluster has been resampled.}. Then, we have mapped the nearest density peak to those previously found. The resulting confidence levels showed in Fig.~\ref{fig:peak.significance} are consistent with the previously found peak positions.

\begin{figure}
\begin{center}
\includegraphics[width=1.0\columnwidth, angle=180]{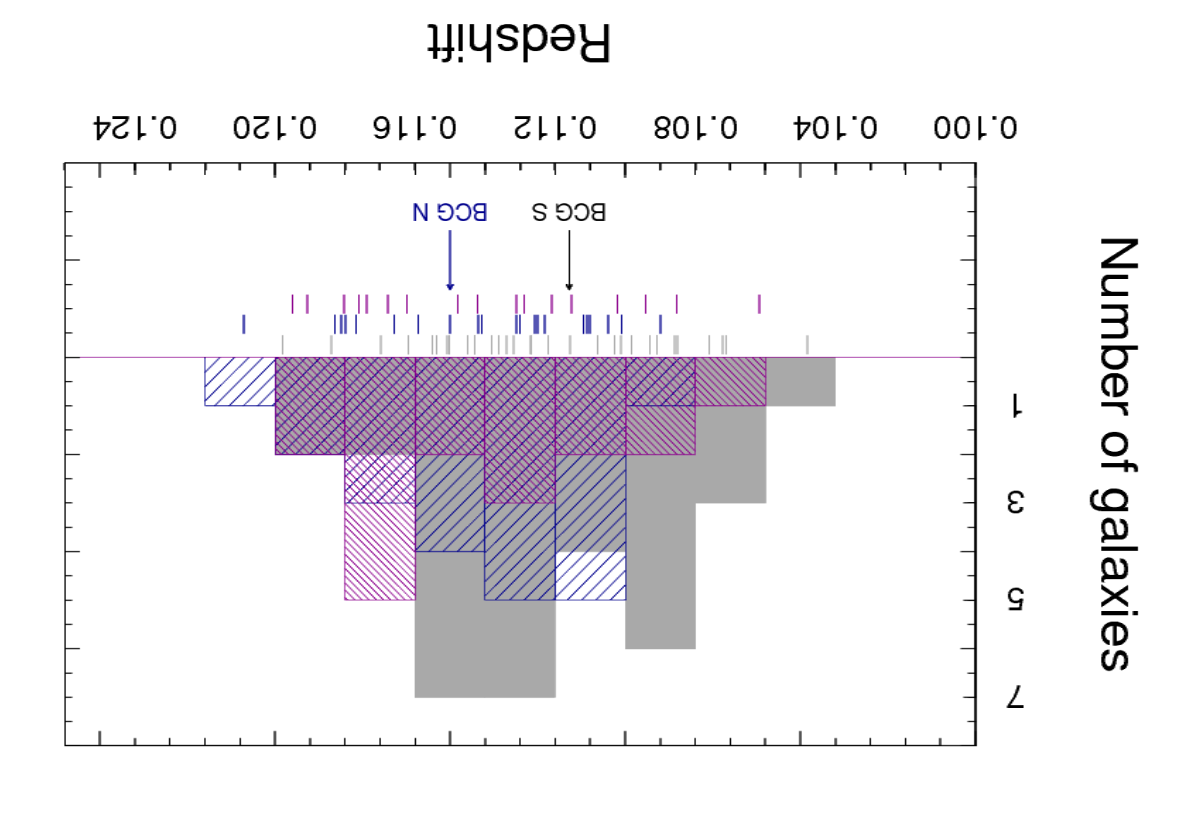}    
\caption{Galaxy classification performed by the {\sc 2d-Mclust} showing the members belonging to A2034S ({\it gray}), A2034N ({\it blue}) and A2034W ({\it purple}). From our spectroscopic sample, only nine galaxies were not classified among the previous groups. The arrows show the redshift of the Southern and Northern BCGs. Dynamically, the groups are characterized as $\bar{z}_S=0.1122\pm0.0006$ and $\sigma_v/(1+z)=958$ km s$^{-1}$, $\bar{z}_N=0.1141\pm0.0007$ and $\sigma_v/(1+z)=873$ km s$^{-1}$ and $\bar{z}_W=0.1140\pm0.0001$ and $\sigma_v/(1+z)=1061$ km s$^{-1}$. The hypothesis of Gaussianity cannot be ruled out within 99\% c.l. according to the Anderson-Darling test.}
\label{fig:2d.histogramas}
\end{center}
\end{figure}

To keep our focus on the main structures we have selected only the galaxies located in the densest part of the field. Then, we have applied the $n$-dimensional Gaussian mixture model algorithm {\sc Mclust} \citep{mclust} to find the best classification of those galaxies. The preferred model with 3 groups is very strongly favoured in relation to the 4-group ($\Delta{\rm BIC}=10$) or even the 5-group ($\Delta{\rm BIC}=14$) models \citep{kass95}. As we can see in Fig.~\ref{fig:peak.significance}, the assigned groups follow the previous relevant density peaks. This classification was then used as the basis to characterize the dynamical state of each structure from the available spectroscopic members. The results are summarized in Tab.~\ref{fig:2dMclust.resultados}  and presented in  Fig.~\ref{fig:2d.histogramas}. For all recovered groups we cannot discard the null hypothesis of normality within 99\% c.l. according to the Anderson-Darling test. The merging system related groups are separated by $\delta v / (1+\bar{z})=497\pm255$ km s$^{-1}$ (68 \% c.l.) along the line-of-sight.

\begin{table}
\begin{center}
\caption{Dynamical parameters of A2034S,  A2034N and A2034W  recovered by the { \sc 2D-Mclust}. }
\begin{tabular}{lccc}
\hline
\hline
  & A2034S & A2034N & A2034W \\
\hline
Members		 & $32$	       & $21$        & $17$ \\
$\bar{z}$                       & $0.1122(6)$ & $0.1141(7)$ & $0.1140(1)$ \\
$\sigma_v /(1+z)$ (km s$^{-1}$) &  $958$      & $873$       & $1061$\\
\hline
\hline
\end{tabular}
\label{fig:2dMclust.resultados}
\end{center}
\end{table} 

\section{A2034S \& A2034N merger dynamics}
\label{sec:two.body}

Our previous dynamical analysis has confirmed the existence of a third structure in the system A2034. However, since the X-ray data brings no hints that a multi-body collision has been taking place, we will explore the simplest model of a bimodal collision between A2034S and A2034N using the 2-body approach implemented on the {\sc Monte Carlo Merger Analysis Code} \citep[MCMAC;][]{dawson}.  This code requires just simple input parameters, listed on Tab.~\ref{tab:input.dawson}, that were later resampled through $2.5\times10^5$ realizations and applied to the model afterward.

\begin{table}
\begin{center}
\caption{Inputs for the MCMAC code: the masses, subcluster's redshits and the projected separation.}
\begin{tabular}{lccc}
\hline
\hline
Parameter & Value & Uncertainty & Unity \\
\hline
$M_{200}^{\rm S}$  & $2.35$    & $0.99$      & $10^{14}$ M${\odot}$ \\[5pt]
$M_{200}^{\rm N}$  & $1.08$    & $0.71$      & $10^{14}$ M${\odot}$ \\[5pt]
$z_{\rm S}$	   & $0.1122$ & $0.0006$  & --	\\[5pt]
$z_{\rm N}$	   & $0.1141$ & $0.0007$  & --\\[5pt]
$d_{\rm proj}$     &	$727$    & $142$         & kpc\\
\hline
\hline
\end{tabular}
\label{tab:input.dawson}
\end{center}
\end{table}

The MCMAC in its original conception considers the merger axis location equally probable between $0^\circ$ and $90^\circ$ from the plane of the sky. However, our previous experience \citep[][]{Monteiro-Oliveira17a,Monteiro-Oliveira17b} have shown that we can constraint these quantity based both on the measured shock projected velocity $v_{\rm shock}$ and the line-of-sight separation $\delta v / (1+\bar{z})$. Given the shock estimation measured by  \cite{owers14} ($v_{\rm shock}=2057$ km s$^{-1}$), we have estimated the perpendicular relative subcluster's velocity 
as $1000\pm500$ km s$^{-1}$ since it corresponds only to a fraction of the shock velocity \citep[e.g.][]{Machado+2015}. These procedures lead to a merger axis located at $\alpha=27^\circ \pm14^\circ$   with respect to the plane of the sky. As an input for MCMAC,  we have adopted a prior allowing $\alpha$ to assume any value between  $0^\circ$ and $35^\circ$ with equal probability. The MCMAC results are presented in Tab.~\ref{tab:output.dawson}.

\begin{table}
\begin{center}
\caption[]{The MCMAC output parameters. $\alpha$ is the merger angle, $d_{3D}(t_{\rm  obs})$ is the current 3D distance, $d_{3D_{\rm  max}}$ is the distance at apoapsis, $v_{3D}(t_{\rm  obs})$ is the 3D observed velocity, $v_{3D}(t_{\rm  col})$ is the 3D velocity at collision time, $v_{3D_{\rm max}}$ is the free fall velocity, $TSC_0$ is the time since the pericentric passage for a outcoming system, $TSC_1$ is the time since the pericentric passage for a incoming system and $T$ is the period between two successive collisions.}
\begin{tabular}{lccccccc}
\hline
\hline
 \multicolumn{2}{c}{Parameter} & Median & 68 \% c.l.  \\
\hline
$\alpha$	       & degree          &  23   & 19 -- 35 \\
$d_{3D}(t_{\rm obs})$     &(Mpc )         &  0.80 & 0.65 --  0.94\\
$d_{3D_{\rm max}}$        & (Mpc)         & 1.43 & 0.64 -- 2.07\\
$v_{3D}(t_{\rm  obs})$     & (km s$^{-1}$) & 1063 & 567 -- 1610 \\
$v_{3D}(t_{\rm  col})$     & (km s$^{-1}$) & 1767 & 1433 -- 2072\\
$v_{3D_{\rm max}}$    & (km s$^{-1}$) & 2331 & 2110 -- 2591 \\
$TSC_0$               & (Gyr)        & 0.56 & 0.34 -- 0.71\\
$TSC_1$               & (Gyr)        & 2.99 & 0.88 -- 4.97\\
$T$                   & (Gyr)        & 3.58 & 1.68 -- 5.55\\
\hline
\hline
\end{tabular}
\label{tab:output.dawson}
\end{center}
\end{table}

The MCMAC predicts two possible merger scenarios: an outgoing system, i.e., seen after the pericentric passage or an incoming system, where the subclusters have reached the apoapsis and are now going to a new encounter. If the outgoing scenario is true, A2034S\&N are currently seen $0.56_{-0.22}^{+0.15}$ Gyr after the pericentric passage. This age is, within 99\% c.l., compatible with those previously estimated by \cite{owers14} based on the shock position in the X-ray map, indicating that the outgoing scenario is preferred. The model also points that the collision tri-dimensional velocity was $1767_{-334}^{+305}$ km s$^{-1}$, being posteriorly reduced to $1063_{-496}^{+547}$ km s$^{-1}$ at the observation time. Regarding their relative tridimensional position, the subclusters are nearly halfway to the apoapsis, $1.43_{-0.79}^{+0.64}$ Mpc, given their current separation of $0.80_{-0.15}^{+0.14}$ Mpc.

\section{Discussion}
\label{sec:discussion}

\subsection{The merger between A2034S and A2034N}
\label{sec:merger}

We present here the weak lensing analysis of A2034 based on three deep filters which have allowed us to select the galaxy populations (foreground, red members and background) more accurately than previous works \citep{sevenmergers,vanweeren11,owers14}. Regarding the cluster members, both numerical density maps (Fig.~\ref{fig:density}) point to the presence of three main concentrations, being two of them nearly aligned with the North-South direction and centred on bright giant galaxies (BCGs).

The recovered total mass distribution map shows a somewhat crowded field. However, two of those mass clumps can be directly related to each optical subcluster due to their proximity in relation to each BCG. Actually, both BCG positions are coincident with the related mass peaks within 95\% c.l. in A2034S and 68\% c.l. in A2034N. The merging system total mass was evaluated as $3.54_{-1.17}^{+0.95}\times10^{14}$ M$_\odot$ giving a mass ratio of $M_S/M_N=2.2_{-1.7}^{+1.1}$, characterizing the system as a major merger or even a semi-major merger considering the error bars \citep{martel14}. According to our model, A2034S appears as the most massive,  with $M_{200}^S=2.35_{-0.99}^{+0.84}\times 10^{14}$ M${\odot}$, whereas A2034N has $M_{200}^N=1.08_{-0.71}^{+0.51}\times 10^{14}$~M${\odot}$. Both subclusters are separated by a projected distance of  $727_{-142}^{+131}$ kpc.

Our proposed reconstruction for the mass distribution in the merging system A2034S\&N shows a reasonable  agreement with the findings of \cite{sevenmergers} (see their Fig.~9). Their results show the presence of two mass concentrations with similar significance (regions ``C'' and ``S'') close to BCG S. The choice of the mass clump ``C'' as representative for A2034S was done based on its spatial concordance with the X-ray peak whereas the clump S was related to the Southern X-ray emission. However, this bimodal configuration in A2034S could not be reproduced in our data even when we changed the size of the smoothing filter used to make the mass map (Fig.~\ref{fig:mass.map}). The same argument is valid for their clumps ``W1''and ``W2'' detected as a single structure in our mass map (\#3). Regarding A2034N, we found similar results showing that the mass concentration appears a little bit Eastern in relation to the BCG N.  It is worth at this point to highlight that our source sample was comprised of 28 background galaxies by arcmin squared, whereas \cite{sevenmergers} had 52.4.

\begin{figure}
\begin{center}
\includegraphics[width=1.0\columnwidth, angle=180]{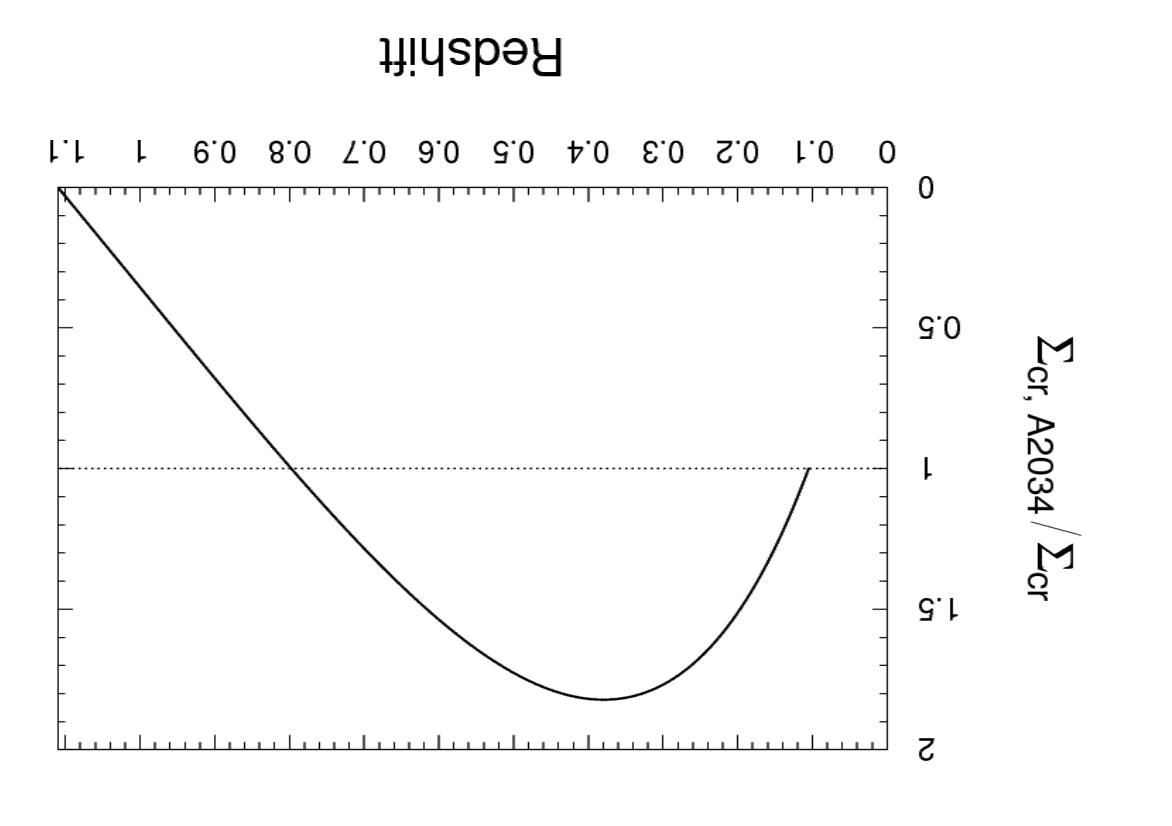}    
\caption{Ratio between the surface critical density at A2034's redshift and at a given redshift ({\it black} continuous line). In this model, we have fixed all background sources at the mean redshift $\bar{z}=1.11$ as found in Sec.~\ref{A2034.tab:masses} after a colour cut in the COSMOS sample. The factor $\xi = \Sigma_{cr,{\rm  A2034}}/\Sigma_{cr}$ allows us to estimate the expected mass at a given redshift by a direct comparison with the values presented in Tab.~\ref{A2034.tab:masses} ($M_{200}(z)=M_{200}(z=0.114)\times\xi^{-1}$). The horizontal {\it dotted} line corresponds to unitary ratio, for a comparison.}
\label{fig:signal}
\end{center}
\end{figure}

Regarding the nature of the surrounding mass clumps (\#5, \#6, \#7 and \#8) we can only speculate. Those structures appear on the mass map but have no optical or X-ray counterparts. Actually, based on the weak-lensing mass reconstruction only we can not even say if those clumps correspond to individual structures or to a collection of small structures on the same line-of-sight  \citep{Liu16}. In Tab.~\ref{A2034.tab:masses} we have estimated their masses assuming that they are at the same redshift as A2034, and those range from 0.6 to 1.8 $\times 10^{14}$M$_\odot$, excluding the peak \#8 whose location is too close to the border region to provide a trustful mass determination. However, given the non-detection of counterparts (Fig.~\ref{fig:optical.counter}), it is reasonable to assume that they could arise from the large scale structure. In that case, those masses can be overestimated by a factor up to 1.8 (at z$\sim$0.4; Fig.~\ref{fig:signal}).

Given the depth of our data, one would imagine that a full investigation on the reality of the lowest S/N mass peaks or a search for the respective counterparts for those structures would probably require space-based imaging. Regarding the mass clumps \#3 and \#4 we are not mentioning them here because given their spatial positions they can be associated with A2034W, which we will discuss in Sec.~\ref{sec:A2034W}.

We can classify the system A2034S\&N as a dissociative merger \citep{dawson} since, within our mass map precision, the X-ray peak related to A2034S is found $169_{-42}^{+48}$ arcsec apart from the mass peak and $91\pm1$ arcsec in relation to the BCG S. In A2034N, as pointed by previous studies \citep{kempner03,owers14}, its gas content was stripped out during the collision in the sense that we cannot identify any gas concentration related to the mass clump. This configuration, therefore, resembles the famous Bullet Cluster \citep[e.g.][]{clowe04,clowe06} where the gas content of both subclusters was displaced in relation to the galaxies and the total mass distribution. The main difference is that, according to our dynamical analysis, the merger axis of A2034S\&N present comparatively a larger angle ($\alpha=27^\circ \pm14^\circ$) in relation to the plane of the sky. Although this estimate has been previously proposed by \cite{owers14}, they found this considering the subclusters' relative velocity to be equal to the shock propagation, which is known to be less precise \citep[e.g.][]{springel07,Machado+2015}. Moreover, the BCGs could not be at rest after the collision which may bias the value of $\delta v$ estimated by only their redshifts.

The distribution of the members with spectroscopic data is well represented by a single Gaussian and this implies that the subclusters are no longer separated along the line-of-sight. In this case, the redshift alone cannot be used as a trustworthy tool to classify the member galaxies according to their host group. So, we turn to the galaxy projected spatial distribution. In fact, our results showed that  A2034S\&N are separated in the line-of-sight by only $\delta v / (1+z)=403\pm228$ km s$^{-1}$.

Departing from the mass posteriors we can apply the scaling relation proposed by \cite{biviano06} to evaluate the expected value of the subcluster velocity dispersion before the collision, under the assumption that it occurred with no mass loss. The comparison with the measured velocity dispersion can be used as an indicator of the dynamical effect of the merger on the galaxies. The pre-merger values are $675_{-96}^{+97}$ km s$^{-1}$ for A2034S and  $548_{-117}^{+122}$ km s$^{-1}$ for A2034N which leads to  boost factors  $f\equiv\sigma_{\rm obs}/\sigma_{\rm pre}$ of  $f_{\rm S}=1.42^{+0.29}_{-0.36}$ and $f_{\rm N}=1.59^{+0.42}_{-0.56}$. These results are another indication that the system is seen less than 1 Gyr after the pericentric passage \cite[see Fig.~29 in][]{pinkney} and suggest that our outgoing scenario is preferred in relation to the incoming configuration. In this case, we found that the system is seen $0.56_{-0.22}^{+0.15}$ Gyr after the pericentric passage.

\subsection{A2034W}
\label{sec:A2034W}

After a complete description of the merging between A2034S and A2034N, which presents a very good match with the observational features, a question yet remains open: what is the nature of A2034W?

\begin{figure}
\begin{center}
\includegraphics[width=\columnwidth,angle=-0]{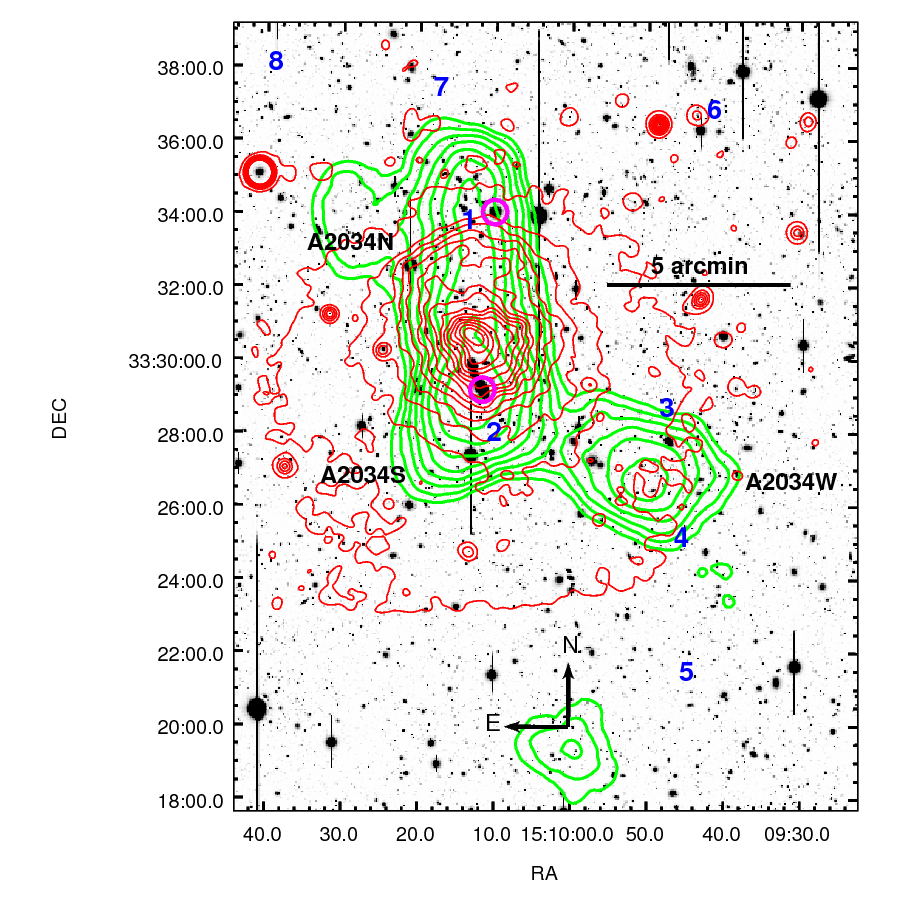} 
\caption{$R_C$ image overlaid with X-ray {\it Chandra} contours ({\it red}) and the spatial numeric distribution of the red cluster members ({\it green}). The numbers are positioned at the mass peak positions. The third galaxy concentration, A2034W, is located nearby the south emission excess present in the X-ray distribution. In spite of the proximity, the correspondence of the galaxy clump A2034W with the nearby mass clumps \#3 and \#4 is not obvious.}  
\label{fig:Xray.galaxies}
\end{center}
\end{figure}

The structure is characterized by $\bar{z}_W=0.1140\pm0.001$ and $\sigma_v/(1+\bar{z})=1061$ km s$^{-1}$ according to our dynamical analysis from 17 identified members with available redshift.
In contrast with the red members spatial distribution, the recovered mass field does not show a clear counterpart for A2034W. In fact, it is located between two mass peaks (\#3 and \#4) as we can see in  Figs.~\ref{fig:boot.peaks} and \ref{fig:Xray.galaxies}. Applying the same approach as used to determine the uncertainty on the A2034S\&N mass peaks centre, we found no spatial coincidence among A2034W and the mass peaks \#3 or \#4 (see Fig.~\ref{fig:boot.peaks}). Therefore, within the limitation of our analysis, the mass counterpart for A2034W remains unclear. The hypothesis that A2034W has no dark matter counterpart although possible is very unlikely.

We found no remarkable galaxy concentration at the position claimed by \cite{owers14} and posteriorly by \cite{Shimwell16} (see their Fig.~3). We warn that this concentration was identified based only on the spectroscopic members. However, the choice of the targets for this kind of observation can induce a bias on their spatial distribution. On the other side, the use of photometric selected red galaxies avoid this effect since their selection is done based only on colour criteria.

The red members distribution closely follow the radio emission map presented by \cite{Shimwell16}. In particular, we found that their region ``C'' (see their Fig.~1) is spatially coincident with the position of A2034W. We also have found a red members concentration around their region F.

Another question we can propose is: what is the relationship between A2034W and the south emission excess in the X-ray distribution? As pointed out by  \cite{kempner03} the properties of this emission are not consistent with an equilibrium state. However, it is cooler than the gas belonging to A2034S\&N to be considered part of the current merger. This is also corroborated by our general analysis of A2034S\&N. A possible hypothesis is that this cool gas could be a remaining part of a merger happened in the past and involving A2034W \citep[e.g.][]{kempner03}.

Finally, our analysis points towards the conclusion of \cite{Shimwell16} that the scenario found in the A2034 field is more complicated than proposed in previous analyses. This complexity can be a hint that previous events had taken place there. For example, a previous merger involving A2034W can explain the south excess in the X-ray distribution, as suggested by \cite{kempner03}. However, further hidrodynamical simulation is required to provide a comprehensive explanation for this complex scenario found in A2034. 

\section{Summary}
\label{sec:summary}

\begin{itemize}

    \item Despite the complexity of the field, the main system is well described by a bimodal merging between A2034N and A2034S closely aligned with the North-South direction. This fact is corroborated by the good match between the recovered age from our two-body analysis and those got by observational X-ray features;
    
    \item We found the individual masses  $M_{200}^S=2.35_{-0.99}^{+0.84}\times 10^{14}$ M${\odot}$ and $M_{200}^N=1.08_{-0.71}^{+0.51}\times 10^{14}$~M${\odot}$, leading to a mass ratio  $M_S/M_N=2.2_{-1.7}^{+1.1}$;
    
    \item The X-ray morphology presents only one peak, related to A2034S. It shows a detachment of $91\pm1$ arcsec and $169_{-42}^{+48}$ arcsec respectively from the BCG S and its related mass clump. On the other side, there is, within 95\% c.l., an agreement between both BCG S and its mass peak positions. Regarding A2034N, its gas content seems to have been also stripped out. 
    
    \item The cluster member classification based on galaxy spatial distribution plus redshift has revealed that A2034S\&N are located not far from each other in relation to the line-of-sight. The relative velocity is only $\delta_v /(1+\bar{z}) = 497\pm255$ km s$^{-1}$; combined with an estimation of the perpendicular velocity based on the shock propagation, we found that the merger axis is located  at $27^\circ \pm14^\circ$  from the plane of the sky;
    
    \item The two-body model predicts that the collision occurred $0.56_{-0.22}^{+0.15}$ Gyr ago with a 3D-velocity of $1767_{-334}^{+305}$ km s$^{-1}$. In spite of the model degeneracy, both shock presence and the measured boost in the velocity dispersion support the outgoing scenario;
    
    \item We confirm the existence of the Western structure A2034W although our analysis does not provide a way to identify its associated dark matter mass clump.
    
\end{itemize}

\section*{Acknowledgements}
\addcontentsline{toc}{section}{Acknowledgements}

The authors thank Cristiano Sousa and Sarah Bridle for their collaboration in earlier phases of this project and Prof. Gast\~ao Lima Neto for providing a detailed description of the {\it Chandra} images and for his helpful contribution with image edition. RMO thanks Tim Shimwell for sharing the radio data and for helpful discussions regarding this very interesting cluster. RMO also thanks the financial support provided by CAPES and CNPq (142219/2013-4) and the project \textit{Casadinho} PROCAD - CNPq/CAPES (552236/2011-0). The authors would also like to acknowledge support from the Brazilian agencies CNPq and FAPESP (\textit{projeto tem\'atico} PI:LSJ 12/00800-4; ESC 2014/13723-3; ALBR 309255/2013-9).

This paper is based in part on observations obtained at the Gemini Observatory, which is operated by the Association of Universities for Research in Astronomy, Inc., under a cooperative agreement with the NSF on behalf of the Gemini partnership: the National Science Foundation (United States), the National Research Council (Canada), CONICYT (Chile), Ministerio de Ciencia, Tecnolog\'{i}a e Innovaci\'{o}n Productiva (Argentina), and Minist\'{e}rio da Ci\^{e}ncia, Tecnologia e Inova\c{c}\~{a}o (Brazil), observing run ID: GN-2010A-Q-22. Based in part on data collected at Subaru Telescope, via the time exchange program between Subaru and the Gemini Observatory. The Subaru Telescope is operated by the National Astronomical Observatory of Japan, observing run ID: GN-2007A-C-21.

This work has made use of the computing facilities of the Laboratory of Astroinformatics (IAG/USP, NAT/Unicsul), whose purchase was made possible by the Brazilian agency FAPESP (grant 2009/54006-4) and the INCT-A.

We made use of the NASA/IPAC Extragalactic Database, which is operated by the Jet Propulsion Laboratory, California Institute of Technology, under contract with NASA.




\bibliographystyle{mnras}
\bibliography{A2034_paper_accepted}


\onecolumn

\appendix
\begin{center}

\section{A2034 catalogue}
\label{ap:catalogue}

  \begin{longtable}{@{}cccccccccc@{}}
    \caption{Catalogue of the heliocentric radial velocities in the A2034 field. In the first three columns we have, respectively the general galaxy ID, the galaxy ID for our Gemini-GMOS sample and the ID of the galaxies found in the NED. The magnitudes $B$, $R_C$ and $z'$, measured inside an aperture of 4.6 arcsec, are presented on the following three columns. In the two last columns are presented the radial velocities and its uncertainty in unities of km s$^{-1}$. Since $\sigma_v$ was not available for the NED galaxies, we have estimate this quantity in 100 km s$^{-1}$. The galaxies highlighted as $^\star$ and $^\diamondsuit$ are respectively the BCGs of the subclusters A2034S and A2034N.}\\
    \hline \hline
     ID & IDg &IDn & $\alpha$ (J2000)& $\delta$ (J2000) & $B$ & $R_C$. & $z'$ & $v$ & $\sigma_v$ \tabularnewline 
    \hline
    \endfirsthead
    
    \multicolumn{9}{c}%
    {{\bfseries \tablename\ \thetable{} -- continued from previous page}} \tabularnewline 
    \hline
     ID & IDm &IDn & $\alpha$ (J2000)& $\delta$ (J2000) & $B$ & $R_C$. & $z'$ & $v$ & $\sigma_v$ \tabularnewline 
   \hline
    \endhead
    
    \multicolumn{9}{c}{{Continued on next page}} \tabularnewline  \hline
    \endfoot  
    
   \hline \hline
    \endlastfoot

1  &1  & --  & 15:09:48.8 & 33:25:39.1 & 21.16  & 19.93  & 19.48  & 35376.  & 100. \\
2  &2  & --  & 15:09:46.1 & 33:25:58.6 & 21.86  & 20.42  & 19.88  & 60918.  & 30. \\
3  &3  & --  & 15:09:55.0 & 33:26:11.8 & 21.09  & 19.39  & 18.82  & 34836.  & 30. \\
4  &4  & --  & 15:09:45.3 & 33:26:35.0 & 21.95  & 20.74  & 20.48  & 104568.  & 60. \\
5  &5  & --  & 15:09:49.2 & 33:26:19.0 & 22.02  & 20.15  & 19.51  & 104478.  & 2458. \\
6  &6  & --  & 15:09:49.4 & 33:26:26.3 & 20.95  & 19.22  & 18.63  & 34416.  & 30. \\
7  &7  & --  & 15:10:14.2 & 33:26:45.4 & 21.87  & 20.73  & 20.36  & 83672.  & 100. \\
8  &8  & --  & 15:10:17.3 & 33:26:43.0 & 23.03  & 21.31  & 20.34  & 37204.  & 60. \\
9  &9  & --  & 15:09:36.9 & 33:26:50.3 & 23.16  & 21.35  & 20.76  & 83552.  & 60. \\
10  &10  & --  & 15:10:15.0 & 33:27:03.2 & 21.49  & 19.83  & 19.24  & 34266.  & 30. \\
11  &11  & --  & 15:09:48.2 & 33:27:07.9 & 21.74  & 20.96  & 20.75  & 62417.  & 60. \\
12  &12  & --  & 15:09:46.0 & 33:27:15.8 & 24.46  & 21.61  & 20.13  & 202060.  & 100. \\
13  &13  & --  & 15:09:46.9 & 33:27:18.6 & 23.1  & 20.70  & 19.91  & 120577.  & 100. \\
14  &14  & --  & 15:09:41.7 & 33:27:03.5 & 20.13  & 18.42  & 17.74  & 32797.  & 30. \\
15  &15  & --  & 15:09:51.2 & 33:27:35.8 & 21.82  & 20.30  & 19.78  & 33847.  & 30. \\
16  &16  & --  & 15:10:06.4 & 33:27:33.1 & 23.72  & 21.25  & 20.16  & 167674.  & 100. \\
17  &17  & --  & 15:09:38.1 & 33:27:28.4 & 23.06  & 21.60  & 21.14  & 36215.  & 30. \\
18  &18  & --  & 15:10:16.8 & 33:27:39.7 & 21.72  & 20.11  & 19.56  & 34596.  & 30. \\
19  &19  & --  & 15:10:02.4 & 33:27:59.9 & 21.31  & 20.16  & 19.77  & 33787.  & 30. \\
20  &20  & --  & 15:09:40.9 & 33:28:00.2 & 22.07  & 20.18  & 19.44  & 65385.  & 30. \\
21  &21  & --  & 15:10:17.4 & 33:27:48.3 & 20.81  & 19.09  & 18.51  & 35915.  & 30. \\
22  &22  & --  & 15:10:05.1 & 33:28:02.2 & 22.24  & 20.62  & 20.07  & 33787.  & 30. \\
23  &23  & --  & 15:09:45.0 & 33:27:48.6 & 21.15  & 19.50  & 18.93  & 33907.  & 30. \\
24  &24  & --  & 15:09:42.2 & 33:28:10.6 & 21.67  & 19.76  & 19.05  & 104508.  & 100. \\
25  &25  & --  & 15:09:48.3 & 33:27:55.0 & 21.41  & 19.35  & 18.63  & 72490.  & 30. \\
26  &26  & --  & 15:10:08.8 & 33:28:14.9 & 22.4  & 20.74  & 20.18  & 33217.  & 60. \\
27  &27  & --  & 15:10:03.7 & 33:27:50.8 & 20.34  & 18.67  & 18.10  & 32917.  & 30. \\
28  &28  & --  & 15:10:11.8 & 33:28:30.2 & 22.47  & 20.92  & 20.38  & 34626.  & 30. \\
29  &29  & --  & 15:10:11.9 & 33:28:24.3 & 21.28  & 19.60  & 19.01  & 33007.  & 30. \\
30  &30  & --  & 15:09:38.5 & 33:28:17.1 & 20.67  & 19.54  & 19.10  & 25872.  & 30. \\
31  &31  & --  & 15:09:53.5 & 33:28:25.1 & 20.95  & 19.27  & 18.72  & 35016.  & 30. \\
32  &32  & --  & 15:09:47.1 & 33:28:52.0 & 23.3  & 20.99  & 20.30  & 118688.  & 100. \\
33  &33  & --  & 15:09:43.2 & 33:28:39.4 & 22.94  & 20.45  & 19.66  & 120876.  & 100. \\
34  &34  & --  & 15:09:51.1 & 33:28:35.6 & 21.13  & 19.24  & 18.56  & 84032.  & 30. \\
35  &35  & --  & 15:10:09.2 & 33:28:20.1 & 20.13  & 18.47  & 17.94  & 32138.  & 30. \\
36  &36  & --  & 15:09:50.1 & 33:29:06.1 & 22.61  & 21.02  & 20.51  & 33427.  & 60. \\
37  &37  & --  & 15:10:05.0 & 33:28:54.1 & 19.92  & 18.24  & 17.56  & 33637.  & 30. \\
38  &38  & --  & 15:10:12.2 & 33:28:45.8 & 21.89  & 20.34  & 19.76  & 32258.  & 60. \\
39  &39  & --  & 15:09:41.3 & 33:29:18.1 & 21.39  & 19.66  & 19.07  & 63316.  & 30. \\
40  &40  & --  & 15:09:52.0 & 33:29:38.0 & 21.8  & 20.92  & 20.65  & 16279.  & 100. \\
41  &41  & --  & 15:10:10.1 & 33:29:12.7 & 20.4  & 18.62  & 17.92  & 34116.  & 30. \\
42  &42  & --  & 15:10:10.7 & 33:29:24.4 & 20.24  & 18.55  & 17.90  & 34326.  & 30. \\
43  &43  & --  & 15:09:55.3 & 33:30:12.2 & 22.01  & 20.51  & 20.14  & 32767.  & 30. \\
44  &44  & --  & 15:10:18.4 & 33:29:41.0 & 20.9  & 19.09  & 18.45  & 33996.  & 30. \\
45  &45  & --  & 15:09:36.0 & 33:29:56.2 & 20.73  & 19.07  & 18.51  & 35915.  & 30. \\
46  &46  & --  & 15:10:12.7 & 33:29:33.0 & 19.15  & 17.51  & 16.66  & 34266.  & 30. \\
47  &47  & --  & 15:10:14.9 & 33:29:42.0 & 20.02  & 18.28  & 17.60  & 34476.  & 30. \\
48  &48  & --  & 15:10:07.6 & 33:30:27.9 & 22.53  & 21.00  & 20.38  & 160479.  & 100. \\
49  &49  & --  & 15:09:50.9 & 33:30:32.4 & 23.93  & 21.17  & 20.09  & 160329.  & 100. \\
50  &50  & --  & 15:09:55.4 & 33:30:22.9 & 21.08  & 19.40  & 18.82  & 35076.  & 30. \\
51  &51  & --  & 15:10:25.1 & 33:30:36.0 & 23.36  & 20.94  & 20.21  & 160329.  & 100. \\
52  &52  & --  & 15:10:15.3 & 33:30:18.3 & 20.59  & 18.76  & 18.10  & 34056.  & 30. \\
53  &53  & --  & 15:09:45.9 & 33:30:44.9 & 21.45  & 19.83  & 19.18  & 83762.  & 90. \\
54  &54  & --  & 15:10:10.7 & 33:30:43.6 & 22.88  & 20.98  & 20.23  & 152744.  & 100. \\
55  &55  & --  & 15:09:51.0 & 33:30:50.2 & 20.77  & 19.53  & 19.02  & 63466.  & 100. \\
56  &56  & --  & 15:10:23.6 & 33:30:52.2 & 24.21  & 21.59  & 20.52  & 167914.  & 100. \\
57  &57  & --  & 15:10:08.9 & 33:30:53.0 & 21.21  & 19.58  & 19.03  & 31418.  & 60. \\
58  &58  & --  & 15:10:16.4 & 33:31:31.2 & 22.45  & 20.79  & 20.24  & 35405.  & 30. \\
59  &59  & --  & 15:10:14.0 & 33:31:19.6 & 21.06  & 19.30  & 18.70  & 34506.  & 30. \\
60  &60  & --  & 15:10:17.0 & 33:31:35.7 & 22.84  & 21.31  & 20.80  & 33907.  & 60. \\
61  &61  & --  & 15:10:25.2 & 33:31:30.6 & 23.04  & 20.54  & 19.80  & 116739.  & 100. \\
62  &62  & --  & 15:10:12.9 & 33:31:25.3 & 20.87  & 19.04  & 18.42  & 34836.  & 30. \\
63  &63  & --  & 15:10:12.0 & 33:31:54.0 & 20.64  & 18.84  & 18.21  & 34746.  & 30. \\
64  &64  & --  & 15:10:24.3 & 33:32:32.0 & 22.16  & 20.35  & 19.69  & 63736.  & 30. \\
65  &65  & --  & 15:10:28.0 & 33:32:34.7 & 22.82  & 21.20  & 20.58  & 116979.  & 180. \\
66  &66  & --  & 15:10:16.0 & 33:32:43.6 & 20.74  & 18.96  & 18.34  & 36245.  & 30. \\
67  &67  & --  & 15:10:16.7 & 33:33:15.8 & 21.28  & 19.53  & 18.94  & 33307.  & 30. \\
68  &68  & --  & 15:10:18.0 & 33:33:04.3 & 20.48  & 18.62  & 17.96  & 35465.  & 30. \\
69  &69  & --  & 15:10:11.3 & 33:33:31.6 & 23.39  & 21.09  & 20.33  & 146748.  & 180. \\
70  &70  & --  & 15:10:09.1 & 33:33:11.6 & 21.42  & 19.33  & 18.62  & 64305.  & 60. \\
71  &71  & --  & 15:10:12.8 & 33:33:15.3 & 22.16  & 20.48  & 19.92  & 33667.  & 60. \\
72  &72  & --  & 15:10:31.4 & 33:33:18.6 & 22.27  & 19.86  & 18.99  & 83732.  & 60. \\
73  &73  & --  & 15:10:22.5 & 33:33:32.0 & 22.56  & 20.64  & 19.94  & 64276.  & 60. \\
74  &74  & --  & 15:10:30.2 & 33:33:35.8 & 23.15  & 21.53  & 20.99  & 116979.  & 180. \\
75  &75  & --  & 15:10:17.7 & 33:33:59.2 & 20.38  & 19.44  & 19.13  & 21315.  & 100. \\
76  &76  & --  & 15:10:09.6 & 33:34:05.3 & 21.13  & 19.28  & 18.70  & 34236.  & 30. \\
77  &77  & --  & 15:10:10.7 & 33:34:04.0 & 20.33  & 18.49  & 17.92  & 33127.  & 30. \\
78  &78  & --  & 15:10:13.3 & 33:33:45.0 & 19.93  & 18.05  & 17.31  & 33757.  & 30. \\
79  &79  & --  & 15:10:25.7 & 33:34:01.5 & 22.13  & 20.47  & 19.92  & 49945.  & 90. \\
80  &80  & --  & 15:10:14.4 & 33:34:07.8 & 19.46  & 17.73  & 16.94  & 33337.  & 30. \\
81  &81  & --  & 15:10:20. & 33:34:32.5 & 20.06  & 18.38  & 17.74  & 33277.  & 30. \\
82  &82  & --  & 15:10:15.0 & 33:34:54.6 & 21.02  & 19.16  & 18.50  & 63406.  & 30. \\
83  &83  & --  & 15:10:14.1 & 33:35:17.9 & 21.62  & 19.84  & 19.22  & 35286.  & 60. \\
84  &84  & --  & 15:10:08.0 & 33:35:29.0 & 24.21  & 21.40  & 20.28  & 161738.  & 180. \\
85  &85  & --  & 15:10:07.2 & 33:35:40.5 & 23.45  & 20.82  & 19.71  & 161768.  & 180. \\
86  &86  & --  & 15:10:17.4 & 33:35:32.4 & 21.44  & 19.76  & 19.17  & 33007.  & 60. \\
87  &87  & --  & 15:10:19.4 & 33:35:23.4 & 20.59  & 18.43  & 17.70  & 64006.  & 30. \\
88  &88  & --  & 15:10:16.0 & 33:36:03.3 & 20.83  & 19.08  & 18.47  & 34956.  & 30. \\
89  &89  & --  & 15:10:27.1 & 33:35:52.0 & 20.36  & 18.53  & 17.85  & 34056.  & 30. \\
90  &90  & --  & 15:10:24.3 & 33:36:22.3 & 20.86  & 19.12  & 18.52  & 33037.  & 30. \\
91  &91  & --  & 15:10:20.3 & 33:36:50.7 & 20.79  & 19.05  & 18.44  & 34056.  & 30. \\
92  &--  & 1  & 15:10:11.4 & 33:29:54.3 & 19.68  & 18.05  & 17.30  & 32527.  & 100. \\
93  &--  & 2  & 15:10:14.6 & 33:31:49.5 & 19.9  & 18.17  & 17.50  & 33727.  & 100. \\
94  &--  & 4  & 15:10:19.2 & 33:30:45.4 & 20.28  & 18.57  & 17.97  & 33937.  & 100. \\
95$^\star$  &--  & 5  & 15:10:11.7 & 33:29:10.4 & 19.21  & 17.65  & 16.69  & 33457.  & 100. \\
96  &--  & 6  & 15:10:00.8 & 33:30:50.4 & 19.82  & 18.04  & 17.28  & 32707.  & 100. \\
97  &--  & 7  & 15:10:20.3 & 33:29:25.4 & 19.56  & 17.90  & 17.12  & 33067.  & 100. \\
98  &--  & 8  & 15:10:02.2 & 33:32:16.8 & 20.02  & 18.23  & 17.53  & 32677.  & 100. \\
99  &--  & 9  & 15:10:20.2 & 33:29:09.5 & 18.73  & 17.36  & 16.35  & 32557.  & 100. \\
100  &--  & 10  & 15:10:12.0 & 33:33:22.5 & 19.94  & 18.17  & 17.50  & 35376.  & 100. \\
101  &--  & 11  & 15:10:24.8 & 33:30:16.6 & 20.40  & 18.36  & 18.02  & 55012.  & 100. \\
102  &--  & 12  & 15:10:04.1 & 33:28:23.5 & 19.65  & 18.03  & 17.33  & 32707.  & 100. \\
103$^\diamondsuit$  &--  & 13  & 15:10:10.2 & 33:34:01.5 & 19.18  & 17.47  & 16.64  & 34476.  & 100. \\
104  &--  & 14  & 15:10:21.4 & 33:28:14.0 & 21.08  & 18.69  & 17.95  & 104088.  & 100. \\
105  &--  & 15  & 15:10:12.0 & 33:34:15.3 & 20.98  & 19.08  & 18.41  & 64425.  & 100. \\
106  &--  & 16  & 15:10:11.7 & 33:27:30.3 & 20.46  & 18.74  & 18.11  & 32108.  & 100. \\
107  &--  & 17  & 15:09:58.6 & 33:28:25.1 & 19.81  & 18.12  & 17.54  & 32527.  & 100. \\
108  &--  & 18  & 15:09:55.3 & 33:29:11.2 & 19.91  & 18.17  & 17.46  & 35196.  & 100. \\
109  &--  & 19  & 15:09:59.8 & 33:27:46.2 & 19.05  & 17.57  & 16.70  & 33037.  & 100. \\
110  &--  & 20  & 15:10:12.8 & 33:34:59.2 & 20.16  & 18.33  & 17.66  & 33877.  & 100. \\
111  &--  & 21  & 15:10:18.4 & 33:26:54.2 & 19.55  & 17.94  & 17.21  & 35495.  & 100. \\
112  &--  & 23  & 15:10:16.8 & 33:35:04.4 & 21.5  & 19.44  & 18.73  & 64246.  & 100. \\
113  &--  & 24  & 15:10:07.3 & 33:35:16.9 & 20.77  & 18.76  & 18.08  & 64395.  & 100. \\
114  &--  & 25  & 15:09:57.4 & 33:27:14.9 & 19.22  & 17.53  & 16.67  & 35256.  & 100. \\
115  &--  & 27  & 15:09:56.3 & 33:27:08.0 & 19.71  & 17.98  & 17.23  & 35705.  & 100. \\
116  &--  & 28  & 15:09:52.2 & 33:34:04.7 & 22.54  & 19.93  & 19.08  & 112422.  & 100. \\
117  &--  & 29  & 15:10:27.9 & 33:34:51.1 & 19.98  & 18.18  & 17.49  & 34206.  & 100. \\
118  &--  & 30  & 15:10:33.4 & 33:33:30.8 & 19.08  & 17.70  & 17.06  & 14750.  & 100. \\
119  &--  & 31  & 15:10:21.3 & 33:26:00.5 & 19.51  & 17.83  & 17.08  & 34956.  & 100. \\
120  &--  & 32  & 15:09:50.4 & 33:34:16.1 & 19.74  & 18.11  & 17.55  & 32378.  & 100. \\
121  &--  & 33  & 15:10:21.4 & 33:36:08.8 & 20.49  & 18.74  & 18.05  & 64066.  & 100. \\
122  &--  & 34  & 15:09:58.8 & 33:25:48.1 & 19.07  & 17.55  & 16.70  & 35376.  & 100. \\
123  &--  & 36  & 15:09:53.7 & 33:26:16.7 & 19.77  & 18.11  & 17.40  & 31838.  & 100. \\
124  &--  & 37  & 15:09:47.3 & 33:27:47.0 & 19.42  & 17.82  & 17.11  & 35825.  & 100. \\
125  &--  & 38  & 15:10:41.2 & 33:31:04.4 & 21.47  & 18.98  & 18.15  & 83372.  & 100. \\
126  &--  & 39  & 15:09:46.6 & 33:34:27.5 & 21.09  & 19.54  & 19.02  & 33847.  & 100. \\
127  &--  & 40  & 15:09:43.2 & 33:28:33.0 & 22.73  & 20.10  & 19.21  & 132208.  & 100. \\
128  &--  & 41  & 15:09:54.6 & 33:36:23.2 & 20.26  & 19.27  & 18.90  & 32677.  & 100. \\
129  &--  & 42  & 15:09:40.2 & 33:30:40.2 & 19.18  & 17.58  & 16.79  & 35765.  & 100. \\
130  &--  & 43  & 15:10:43.5 & 33:30:37.6 & 19.72  & 17.93  & 17.12  & 35885.  & 100. \\
131  &--  & 44  & 15:10:23.3 & 33:24:37.6 & 20.07  & 18.36  & 17.78  & 32527.  & 100. \\
132  &--  & 45  & 15:10:00.3 & 33:24:14.7 & 20.19  & 18.40  & 17.76  & 35645.  & 100. \\
133  &--  & 46  & 15:10:06.2 & 33:23:46.6 & 20.95  & 19.29  & 18.71  & 33277.  & 100. \\
134  &--  & 47  & 15:10:01.6 & 33:23:59.3 & 19.71  & 18.00  & 17.27  & 33277.  & 100. \\
135  &--  & 48  & 15:09:43.3 & 33:26:44.8 & 19.93  & 18.21  & 17.54  & 33607.  & 100. \\
136  &--  & 49  & 15:10:00.0 & 33:24:02.7 & 20.45  & 18.61  & 17.92  & 35076.  & 100. \\
137  &--  & 50  & 15:10:21.3 & 33:37:57.7 & 20.10  & 18.33  & 17.62  & 35915.  & 100. \\
138  &--  & 51  & 15:10:41.1 & 33:35:04.9 & 19.33  & 17.67  & 16.78  & 34236.  & 100. \\
139  &--  & 52  & 15:10:41.1 & 33:35:04.9 & 19.33  & 17.67  & 16.78  & 34236.  & 100. \\
140  &--  & 53  & 15:09:52.3 & 33:24:38.0 & 20.47  & 18.72  & 18.09  & 34236.  & 100. \\
141  &--  & 54  & 15:10:15.0 & 33:23:14.5 & 19.64  & 17.99  & 17.28  & 33217.  & 100. \\
142  &--  & 55  & 15:10:37.5 & 33:36:35.3 & 19.55  & 17.81  & 16.99  & 32647.  & 100. \\
143  &--  & 56  & 15:09:37.0 & 33:27:37.3 & 19.51  & 17.88  & 17.18  & 35855.  & 100. \\
144  &--  & 57  & 15:09:38.2 & 33:26:54.6 & 19.55  & 18.14  & 17.41  & 31328.  & 100. \\
145  &--  & 58  & 15:10:48.7 & 33:33:33.4 & 20.88  & 19.07  & 18.41  & 33607.  & 100. \\
146  &--  & 59  & 15:10:39.7 & 33:36:52.3 & 18.98  & 17.64  & 16.87  & 13041.  & 100. \\
147  &--  & 60  & 15:10:23.7 & 33:38:52.0 & 19.62  & 17.81  & 17.17  & 35735.  & 100. \\
148  &--  & 61  & 15:09:29.7 & 33:29:09.7 & 20.2  & 18.40 & 17.70  & 34566.  & 100. \\
149  &--  & 62  & 15:09:30.1 & 33:33:26.2 & 20.38  & 18.40  & 17.73  & 64096.  & 100. \\
150  &--  & 63  & 15:09:35.3 & 33:25:54.2 & 19.92  & 18.16  & 17.46  & 33427.  & 100. \\
151  &--  & 64  & 15:09:44.6 & 33:38:01.2 & 19.69  & 17.90  & 17.18  & 32378.  & 100. \\
152  &--  & 65  & 15:09:32.0 & 33:26:51.8 & 20.90  & 18.89  & 18.22  & 59479.  & 100. \\
153  &--  & 66  & 15:10:24.2 & 33:21:54.5 & 19.71  & 18.04  & 17.33  & 34446.  & 100. \\
154  &--  & 67  & 15:09:57.3 & 33:22:00.6 & 19.70  & 18.03  & 17.33  & 33637.  & 100. \\
155  &--  & 68  & 15:10:58.1 & 33:30:37.6 & 19.62  & 17.89  & 17.07  & 34086.  & 100. \\
156  &--  & 69  & 15:09:41.3 & 33:23:17.1 & 20.52  & 18.81  & 18.21  & 33697.  & 100. \\
157  &--  & 70  & 15:09:23.7 & 33:30:05.9 & 19.77  & 18.17  & 17.51  & 33307.  & 100. \\
158  &--  & 71  & 15:09:23.5 & 33:30:25.7 & 19.99  & 18.16  & 17.47  & 42990.  & 100. \\
159  &--  & 72  & 15:09:50.0 & 33:40:06.3 & 19.84  & 18.21  & 17.58  & 34926.  & 100. \\
160  &--  & 73  & 15:10:52.1 & 33:36:57.0 & 20.19  & 18.32  & 17.62  & 34896.  & 100. \\
161  &--  & 74  & 15:09:28.3 & 33:25:09.5 & 19.87  & 18.16  & 17.45  & 35076.  & 100. \\
162  &--  & 75  & 15:09:30.9 & 33:24:14.3 & 19.84  & 18.17  & 17.46  & 35256.  & 100. \\
163  &--  & 76  & 15:09:20.7 & 33:33:52.2 & 19.76  & 18.10  & 17.35  & 32707.  & 100. \\
164  &--  & 77  & 15:09:50.2 & 33:41:29.4 & 19.47  & 17.78  & 17.07  & 33397.  & 100. \\
165  &--  & 78  & 15:09:27.2 & 33:23:50.0 & 19.18  & 17.84  & 17.11  & 26112.  & 100. \\
166  &--  & 79  & 15:10:53.5 & 33:38:48.9 & 19.81  & 17.98  & 17.15  & 33427.  & 100. \\
167  &--  & 80  & 15:10:35.7 & 33:19:57.5 & 21.97  & 19.32  & 18.46  & 117788.  & 100. \\
168  &--  & 81  & 15:09:39.9 & 33:20:52.0 & 20.11  & 18.11  & 17.36  & 59659.  & 100. \\
169  &--  & 82  & 15:10:13.9 & 33:18:40.6 & 19.92  & 18.22  & 17.54  & 35346.  & 100. \\
170  &--  & 83  & 15:09:33.1 & 33:21:12.3 & 18.89  & 17.52  & 16.61  & 26082.  & 100. \\
171  &--  & 84  & 15:09:46.6 & 33:18:10.4 & 19.57  & 17.99  & 17.32  & 31898.  & 100. \\
172  &--  & 85  & 15:09:36.1 & 33:19:15.0 & 20.62  & 18.77  & 18.14  & 59689.  & 100. \\
173  &--  & 86  & 15:11:00.4 & 33:40:20.0 & 20.10  & 18.21  & 17.44  & 33367.  & 100. \\
174  &--  & 87  & 15:11:19.3 & 33:29:59.0 & 19.50  & 17.58  & 16.56  & 35825.  & 100. \\
175  &--  & 88  & 15:09:11.0 & 33:37:39.4 & 19.64  & 18.05  & 17.41  & 34956.  & 100. \\
176  &--  & 89  & 15:09:25.7 & 33:19:52.7 & 19.78  & 18.55  & 18.07  & 42840.  & 100. \\
177  &--  & 90  & 15:09:43.4 & 33:16:52.8 & 19.41  & 17.77  & 17.03  & 34116.  & 100. \\
178  &--  & 91  & 15:10:27.6 & 33:16:00.1 & 19.64  & 17.93  & 17.22  & 34836.  & 100. \\
179  &--  & 92  & 15:09:10.0 & 33:22:19.4 & 19.23  & 17.75  & 17.00  & 33247.  & 100. \\
180  &--  & 93  & 15:10:27.7 & 33:15:23.6 & 20.11  & 18.45  & 17.88  & 33307.  & 100. \\
181  &--  & 94  & 15:09:07.6 & 33:22:12.4 & 20.05  & 18.94  & 18.40  & 33097.  & 100. \\
182  &--  & 95  & 15:10:00.8 & 33:15:05.9 & 19.06  & 17.43  & 16.60  & 33097.  & 100. \\

  \end{longtable} 
  \end{center}


\bsp	
\label{lastpage}
\end{document}